\documentclass[a4paper,11pt]{article}

\usepackage[a4paper,margin=1in]{geometry}
\usepackage{graphicx}
\usepackage{subcaption}
\usepackage[dvipsnames]{xcolor}
\usepackage[colorlinks=true,linkcolor=NavyBlue]{hyperref}
\usepackage{amsmath}
\usepackage{multirow}
\usepackage{physics}
\usepackage{authblk}
\usepackage{indentfirst}
\usepackage{booktabs}
\usepackage{subcaption}
\usepackage{CJKutf8}

\begin{document}
\begin{CJK*}{UTF8}{gbsn}

\title{Study of background from accidental coincidence signals in the PandaX-II experiment}

\author[1]{Abdusalam Abdukerim(阿布都沙拉木·阿布都克力木)}
\author[1]{Wei Chen(陈葳)}
\author[1,2]{Xun Chen(谌勋)\thanks{corresponding author, chenxun@sjtu.edu.cn}}
\author[3]{Yunhua Chen(陈云华)}
\author[4]{Chen Cheng(程晨)}
\author[5]{Xiangyi Cui(崔祥仪)}
\author[6]{Yingjie Fan(樊英杰)}
\author[7]{Deqing Fang(方德清)}
\author[7]{Changbo Fu(符长波)}
\author[8]{Mengting Fu(付孟婷)}
\author[9,10,11]{Lisheng Geng(耿立升)}
\author[1]{Karl Giboni}
\author[1]{Linhui Gu(顾琳慧)}
\author[3]{Xuyuan Guo(郭绪元)}
\author[1]{Ke Han(韩柯)}
\author[1]{Changda He(何昶达)}
\author[1]{Di Huang(黄迪)}
\author[3]{Yan Huang(黄焱)}
\author[12]{Yanlin Huang(黄彦霖)}
\author[1]{Zhou Huang(黄周)}
\author[13]{Xiangdong Ji(季向东)}
\author[14]{Yonglin Ju(巨永林)}
\author[5]{Shuaijie Li(李帅杰)}
\author[15]{Huaxuan Liu(刘华萱)}
\author[1,5,2]{Jianglai Liu(刘江来)}
\author[1]{Wenbo Ma(马文博)}
\author[7,16]{Yugang Ma(马余刚)}
\author[8]{Yajun Mao(冒亚军)}
\author[1,2]{Yue Meng(孟月)}
\author[1]{Kaixiang Ni(倪恺翔)}
\author[3]{Jinhua Ning(宁金华)}
\author[1]{Xuyang Ning(宁旭阳)}
\author[15]{Xiangxiang Ren(任祥祥)}
\author[5]{Changsong Shang(商长松)}
\author[1]{Lin Si(司琳)}
\author[9]{Guofang Shen(申国防)}
\author[15]{Andi Tan(谈安迪)}
\author[14]{Anqing Wang(王安庆)}
\author[16,17]{Hongwei Wang(王宏伟)}
\author[15]{Meng Wang(王萌)}
\author[7]{Qiuhong Wang(王秋红)}
\author[8]{Siguang Wang(王思广)}
\author[4]{Wei Wang(王为)}
\author[14]{Xiuli Wang(王秀丽)}
\author[1,2]{Zhou Wang(王舟)}
\author[4]{Mengmeng Wu(武蒙蒙)}
\author[3]{Shiyong Wu(吴世勇)}
\author[1]{Weihao Wu(邬维浩)}
\author[1]{Jingkai Xia(夏经铠)}
\author[13,18]{Mengjiao Xiao(肖梦姣)}
\author[5]{Pengwei Xie(谢鹏伟)}
\author[1]{Binbin Yan(燕斌斌)}
\author[1]{Jijun Yang(杨继军)}
\author[1]{Yong Yang(杨勇)}
\author[6]{Chunxu Yu(喻纯旭)}
\author[15]{Jumin Yuan(袁鞠敏)}
\author[1]{Ying Yuan(袁影)}
\author[1]{Xinning Zeng(曾鑫宁)}
\author[13]{Dan Zhang(张丹)}
\author[1,2]{Tao Zhang(张涛)}
\author[1,2]{Li Zhao(赵力)}
\author[12]{Qibin Zheng(郑其斌)}
\author[3]{Jifang Zhou(周济芳)}
\author[1]{Ning Zhou(周宁)}
\author[9]{Xiaopeng Zhou(周小朋)}

\affil[1]{School of Physics and Astronomy, Shanghai Jiao Tong University, MOE Key Laboratory for Particle Astrophysics and Cosmology, Shanghai Key Laboratory for Particle Physics and Cosmology, Shanghai 200240, China}
\affil[2]{Shanghai Jiao Tong University Sichuan Research Institute, Chengdu 610213, China}
\affil[3]{Yalong River Hydropower Development Company, Ltd., 288 Shuanglin Road, Chengdu 610051, China}
\affil[4]{School of Physics, Sun Yat-Sen University, Guangzhou 510275, China}
\affil[5]{Tsung-Dao Lee Institute, Shanghai 200240, China}
\affil[6]{School of Physics, Nankai University, Tianjin 300071, China}
\affil[7]{Key Laboratory of Nuclear Physics and Ion-beam Application (MOE), Institute of Modern Physics, Fudan University, Shanghai 200433, China}
\affil[8]{School of Physics, Peking University, Beijing 100871, China}
\affil[9]{School of Physics, Beihang University, Beijing 102206, China}
\affil[10]{Beijing Key Laboratory of Advanced Nuclear Materials and Physics, Beihang University, Beijing, 102206, China}
\affil[11]{School of Physics and Microelectronics, Zhengzhou University, Zhengzhou, Henan 450001, China}
\affil[12]{School of Medical Instrument and Food Engineering, University of Shanghai for Science and Technology, Shanghai 200093, China}
\affil[13]{Department of Physics, University of Maryland, College Park, Maryland 20742, USA}
\affil[14]{School of Mechanical Engineering, Shanghai Jiao Tong University, Shanghai 200240, China}
\affil[15]{School of Physics and Key Laboratory of Particle Physics and Particle Irradiation (MOE), Shandong University, Jinan 250100, China}
\affil[16]{Shanghai Institute of Applied Physics, Chinese Academy of Sciences, Shanghai 201800, China}
\affil[17]{Shanghai Advanced Research Institute, Chinese Academy of Sciences, Shanghai 201210, China}
\affil[18]{Center for High Energy Physics, Peking University, Beijing 100871, China}
\affil[ ]{(PandaX-II Collaboration)}

\maketitle

\abstract{The PandaX-II experiment employed a 580kg liquid xenon
  detector to search for the interactions between dark matter
  particles and the target xenon atoms. The accidental coincidences of
  isolated signals result in a dangerous background which mimic the
  signature of the dark matter. We performed a detailed study on the
  accidental coincidence background in PandaX-II, including the
  possible origin of the isolated signals, the background level and
  corresponding background suppression method. With a
  boosted-decision-tree algorithm, the accidental coincidence
  background is reduced by 70\% in the dark matter signal region, thus
  the sensitivity of dark matter search at PandaX-II is improved.}

keywords: dark matter, xenon detector, background, accidental coincidence, machine learning

\section{Introduction}
\label{sec:intro}
The direct detection of the dark matter particles, especially the
weakly interacting massive particles (WIMPs), is actively carried out
by a couple of experiments all over the world
currently~\cite{NPreviewDirect}. In recent years, the PandaX-II
experiment located in the China Jinping Underground Laboratory
(CJPL)~\cite{NPreviewDirect,Kang:2010zza,Zhao:2020ezy}, which uses the technology
of dual phase liquid xenon time projection chambers (TPCs), has pushed
the limits of cross section between WIMPs and nucleons to a new level
for most of the possible WIMP masses, with other experiments of the
same
type~\cite{Wang:2020coa,Cui:2017nnn,PandaX-II:2016vec,Tan:2016diz,Akerib:2016vxi_LUXNR,Aprile:2018dbl_xenon1tNR,XENON:2017vdw}. The
scattering of incident particles with xenon atoms in the TPC may
produce a prompt scintillation $S1$, which resulted from the
de-excitation of xenon atoms and the recombination process of some
ionized electrons. Some electrons escaping from the recombination may
drift along the electric field inside the TPC and be extracted into
the gaseous region, producing the proportional electroluminescent
scintillation $S2$~\cite{Aprile:2009dv,Aprile:2006kx}. The detected
signals of $S1$ and $S2$ are used to reconstruct the scattering event
in the data analysis. Due to the low probability of scattering events
between WIMPs and ordinary matter, a good physical event requires only
one pair of physically correlated $S1$ and $S2$ within the maximum
electron drift time window inside the TPC. In the last results of the
PandaX-I experiment~\cite{PandaX:2015gpz}, it was realized that the
accidental coincidence of isolated $S1$ and $S2$ within the window
comprises a new type of background, which contributes a number of
events in the signal parameter space to search for
WIMPs. Understanding this type of background and development of
methods to suppress it become important for the improvement of dark
matter detection sensitivity. In the data analysis of PandaX-II with
the full exposure, we made a thorough study of the accidental
background and presented an accurate estimation of its level for all
the three data taken runs, 9, 10 and 11~\cite{Wang:2020coa}.

In this article, we present a detailed introduction on the study of
accidental background in PandaX-II. In Section \ref{sec:tpc}, we
provide a brief introduction to the PandaX-II TPC, the signals and the
backgrounds. Then we discuss the possible origin of the accidental
background in Section \ref{sec:origin}. The estimation of its level is
presented in Section \ref{sec:estimation}. The application of the
boost-decision-tree (BDT) method to suppress the background is given
in Section \ref{sec:bdt}, with the performance presented. At last, we
give a brief summary and outlook in Section \ref{sec:summary}.

\section{TPC, signals and backgrounds of PandaX-II}
\label{sec:tpc}
\begin{figure}[htb]
  \centering \includegraphics[width=0.7\textwidth]{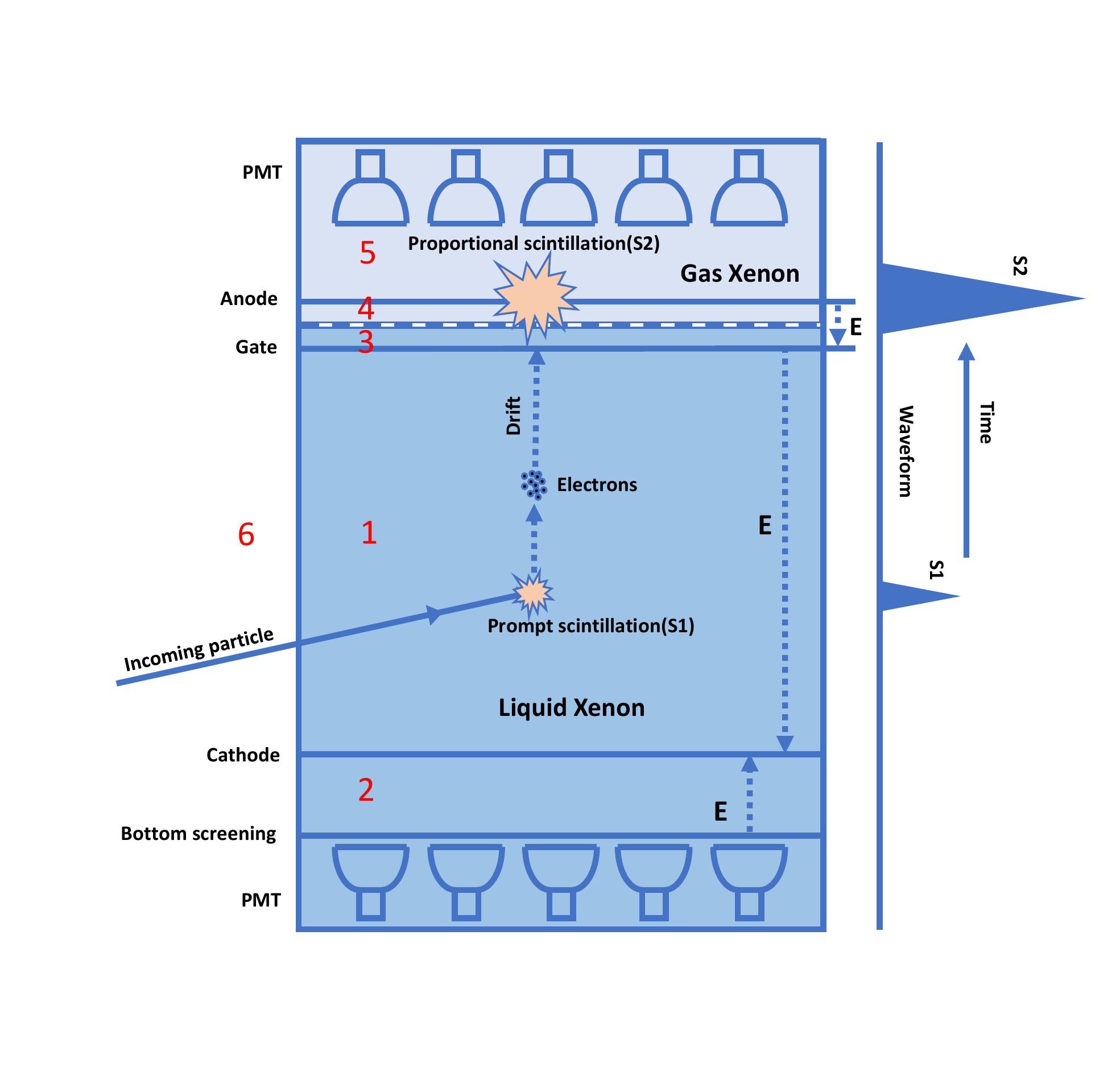}
  \caption{Schematic view of the TPC of PandaX-II, with six regions
    labeled with numbers: 1. the liquid part below the gate and above
    the cathode; 2. the liquid part below the cathode; 3. the liquid
    part above the gate; 4. the gas part below the anode; 5. the gas
    part above the anode; 6. parts outside the inner PTFE walls. A
    recoil event in region 1 may produce $S1$ and $S2$ signals at
    different regions in the detector with a time delay.}
  \label{fig:tpc}
\end{figure}

A detail description of PandaX-II TPC is presented in
Ref.~\cite{Tan:2016diz}. A more detailed schematic view of the TPC is
presented in Figure \ref{fig:tpc}. The near-cylindrical shaped TPC
confined by polytetrafluoroethylene (PTFE) walls, contains both of
gaseous xenon (top) and liquid xenon (bottom) in its
volume. Scintillation light generated inside the TPC is detected by
the two arrays of photo-multiplier tubes(PMTs) located on the top and
bottom region, respectively. The cathode in the bottom part of the TPC
and the gate electrode right below the liquid surface, provide the
drift electric field for ionized electrons and define the sensitive
region of the detector (region 1 in Figure~\ref{fig:tpc}).

Deposited energies by scattering events inside the sensitive region
result in a $S1$ signal, typically with a time spreading\footnote{We
  use the term of ``width'' in following text to represent this
  concept.}  smaller than 150~ns, in very short time, while the
possible $S2$ signal will be produced after a time delay, due to the
limited drift velocity of ionized electrons inside liquid xenon. The
drift velocity of electrons depends only on the electric field
strength, thus the time difference between physically correlated $S1$
and $S2$ can be used to calculate the vertical position of a
scattering event. The maximum drift time for electrons in the
sensitive region is $350\pm8$~$\mu$s in Run 9 and $360\pm8$~$\mu$s in Runs 10 and
11, due to the different drift fields~\cite{Wang:2020coa}. When a ``trigger''
signal exceed the pre-defined threshold is observed during the
ordinary data taking, the digitized waveform of all the PMTs within a
window 500~$\mu$s before and after the trigger time is recorded as an
event. The data processing steps calculate the baseline of each
recorded waveform, search for ``hit'' exceeding a given threshold of
0.25 photoelectrons (PE) and cluster the overlapped hits into
signals. Events of single scattering (with only one $S1$ and $S2$
reconstructed) are selected, and then filtered by the quality cuts to
search for the possible rare scattering from WIMPs.

Recognition, understanding and suppression of the different types of
backgrounds are critical in the data analysis of WIMP searching
experiments because the desired signal rate is very low. In the
PandaX-II experiment, the backgrounds can be categorized into four
types. The electron recoil (ER) backgrounds, mainly from the
radioactive isotopes in the detector material or in the xenon target,
have been studied and understood with the ER calibration data and
Geant4-based Monte Carlo (MC)
simulations~\cite{GEANT4:2002zbu,Allison:2006ve}. The nuclear recoil
(NR) background, mainly from neutrons produced by the ($\alpha$, $n$)
process or spontaneous fission of isotopes in detectors, has been
estimated by the correlated high energy gamma events with the help of
simulation~\cite{Wang:2019opt}. The surface background are created by
daughters of $^{222}$Rn attached on the inner surface of the TPC, with
suppressed $S2$ due to the charge loss on the PTFE wall. The level of
surface background is estimated with a data driven
method~\cite{Wang:2020coa,Zhang:2019evc}. The last one is the
nonphysical accidental background resulted from the falsely pairing of
unrelated $S1$ and $S2$ signals. A large proportion of the accidental
background events have relatively small $S2$ signals, thus are not
easy to be distinguished from the physical NR events (neutron or
WIMPs) by investigating the ratio of $S2/S1$ only. Effective
suppression of the accidental background will improve the discovery
sensitivity of WIMPs greatly.

\section{Origins of the accidental background}
\label{sec:origin}
In the TPC, some $S1$ or $S2$-like signals may be produced solely,
without any other signals from the same source observed by detector.
We call these signals ``isolated''. Since the events with a pair of
$S1$ and $S2$ are used to search for dark matter, the isolated signals
appearing in the same drift window may be paired, resulting in the
accidental background.

\subsection{The isolated $S1$}
\label{sec:isolated_s1}
The origins of isolated $S1$s may be physical or non-physical. The
physical origins might be in the following:
\begin{itemize}
\item tiny sparks on the TPC electrode, no electrons produced;
\item scattering events in the region between the cathode and the
  screening electrode of the bottom array (region 2 in
  Figure~\ref{fig:tpc}), within which no electron could drift into the
  gas xenon, thus no $S2$ could be produced;
\item physical events occur near the bottom wall of the detector with
  all electrons lost due to the imperfect of the drift field, thus
  no $S2$ is produced;
\item scattering events above the anode in the gaseous region (region
  5 in Figure~\ref{fig:tpc}), with no electrons entering the region below
  the anode to produce $S2$;
\item signals produced by single electron extract into the gas region, which are
  mis-identified as $S1$s;
\item possible light leakage from scattering events outside the TPC
  (region 6 in Figure~\ref{fig:tpc}).
\end{itemize}

The dominant non-physical origin of isolated $S1$ is from the dark
noise of the PMT, which produce small hits in the readout waveform of
each PMT. During the event reconstruction, a valid signal should
contain overlapped hits from at least three PMTs. The relatively high
rate of dark noise (average rates are 1.9, 0.17 and 0.23 kHz per PMT
for Runs 9, 10 and 11, respectively) makes it possible for the
formation of small $S1$-like signals by the randomly coincidence of
the dark noises. Since these signals have contribution from the top
PMTs, their top-bottom asymmetry (discussed in section~\ref{sec:vars})
should not be $-1$.

\subsection{The isolated $S2$}
The $S2$ signals are from the electroluminescent of electrons in the
gas region. The isolated $S2$ signals, without exception, resulted
from the same process. The origin of isolated $S2$ can be categorized
into four types:
\begin{itemize}
\item real scattering event in the sensitive region with small energy
  deposition, and the weak $S1$ is not recognized due to the detection
  efficiency;
\item real scattering event in the sensitive region, but it is too
  close to the liquid surface, resulted in overlapped $S1$ and $S2$
  signals, which are recognized as one $S2$;
\item real scattering event in the region above the gate but below the
  anode (region 3 and 4 in Figure~\ref{fig:tpc}), with overlapped $S1$
  and $S2$ signals recognized as one $S2$; the signal may have smaller
  width and asymmetrical shape;
\item the electrons generated with large energy deposition may not be
  extracted into the gas region completely. The rest electrons gather
  on the liquid surface and are released into the gas randomly,
  producing electroluminescent directly.

\end{itemize}

\section{Estimation of accidental background}
\label{sec:estimation}
Since the isolated signals are independent from each other, the level
of accidental background can be calculated by the rates of isolated
$S1$ and $S2$ signals, assuming they follow an uniform distribution
along time in a selected period with same run conditions. Estimation
of the rates of these signals becomes important in this study.

For each of the data taking run, the average rates
$\bar{r}_1$ for isolated $S1$ and $\bar{r}_2$ for
isolated $S2$ are computed by the time weighted average of the corresponding rates:
\begin{align}
  \label{eq:rate_average}
  \bar{r}_1 = \frac{1}{\sum_i T_i}\sum_i r_{1i}\cdot T_i,\\
  \bar{r}_2 = \frac{1}{\sum_i T_i}\sum_i r_{2i}\cdot T_i, 
\end{align}
where $T_i$, $r_{1i}$, $r_{2i}$ are the duration, rates of isolated
$S1$ and $S2$ for each selected period $i$, respectively. The
uncertainties of the rates are calculated as the unbiased standard
errors of the mean value.

\subsection{Tagging of isolated $S1$}
\label{sec:rate_s1}
To calculate the rate of isolated $S1$, we need to recognize this type
of signal correctly in the data. Three methods have been developed to
search for the isolated $S1$ within the range of $(3, 100)$~PE, which
covers the energy region of searching for dark matter. One is based on
a special type of ``random trigger'' data set, with the event
triggered by hardware randomly. The other two methods are based on the
dark matter search data. We describe all of these three methods here.

\subsubsection{Method 1}
\label{sec:m1_s1}
This method is to search isolated $S1$ events in the random trigger
data. The events should satisfy all the required data quality cuts
mentioned in Ref.~\cite{Wang:2020coa}. The rate $r_1$ in one run can
be calculated easily by~\footnote{The subscript ``$i$'' is omitted
  in following formulas.}
\begin{equation}
  \label{eq:is1_rate_m1}
  r_1 = \frac{n_{iS1}}{T},
\end{equation}
where $n_{iS1}$ is the number of qualified isolated $S1$, and $T$ is
the live time of the random trigger events. The method is unbiased,
and is used to estimate the accidental background level in Run
10~\cite{Cui:2017nnn}. Due to the short time of data taking with
random trigger, the long term evolution of the rate can not be
extracted. No random trigger data taking was performed in Run 9, so
this method can only work in Runs 10 and 11.

\subsubsection{Method 2}
\label{sec:m1_s2}
\begin{figure}[hbt]
  \centering
  \includegraphics[width=0.8\textwidth]{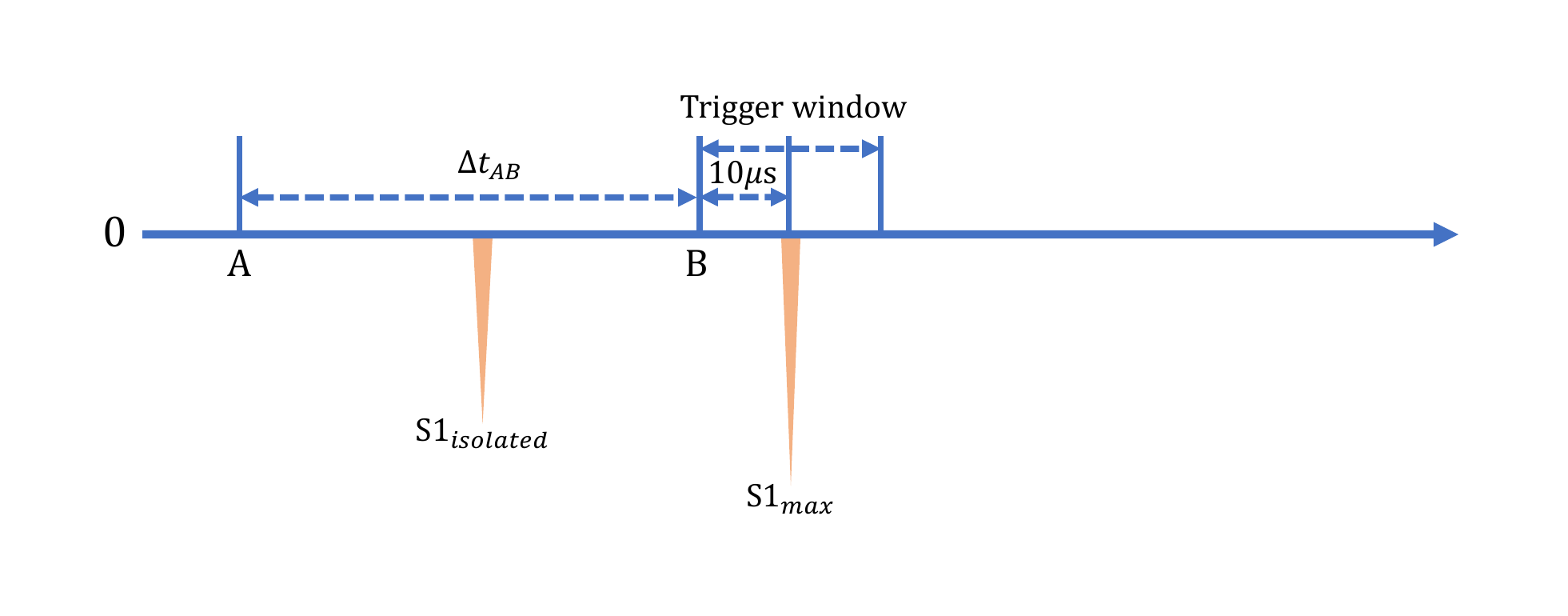}
  \caption{Schematic view on the search of isolated $S1$ in events
    triggered by unpaired $S1$ ($S1_{max}$) in Runs 10 and 11. The
    event has a fixed time window of 1~ms, and the trigger windows is
    within $(490, 510)~\mu$s. The symbol of ``A'' and ``B'' indicates
    the searching window for isolated $S1$. }
  \label{fig:s1_m2}
\end{figure}

\begin{figure}[hbt]
  \centering
  \includegraphics[width=0.65\textwidth]{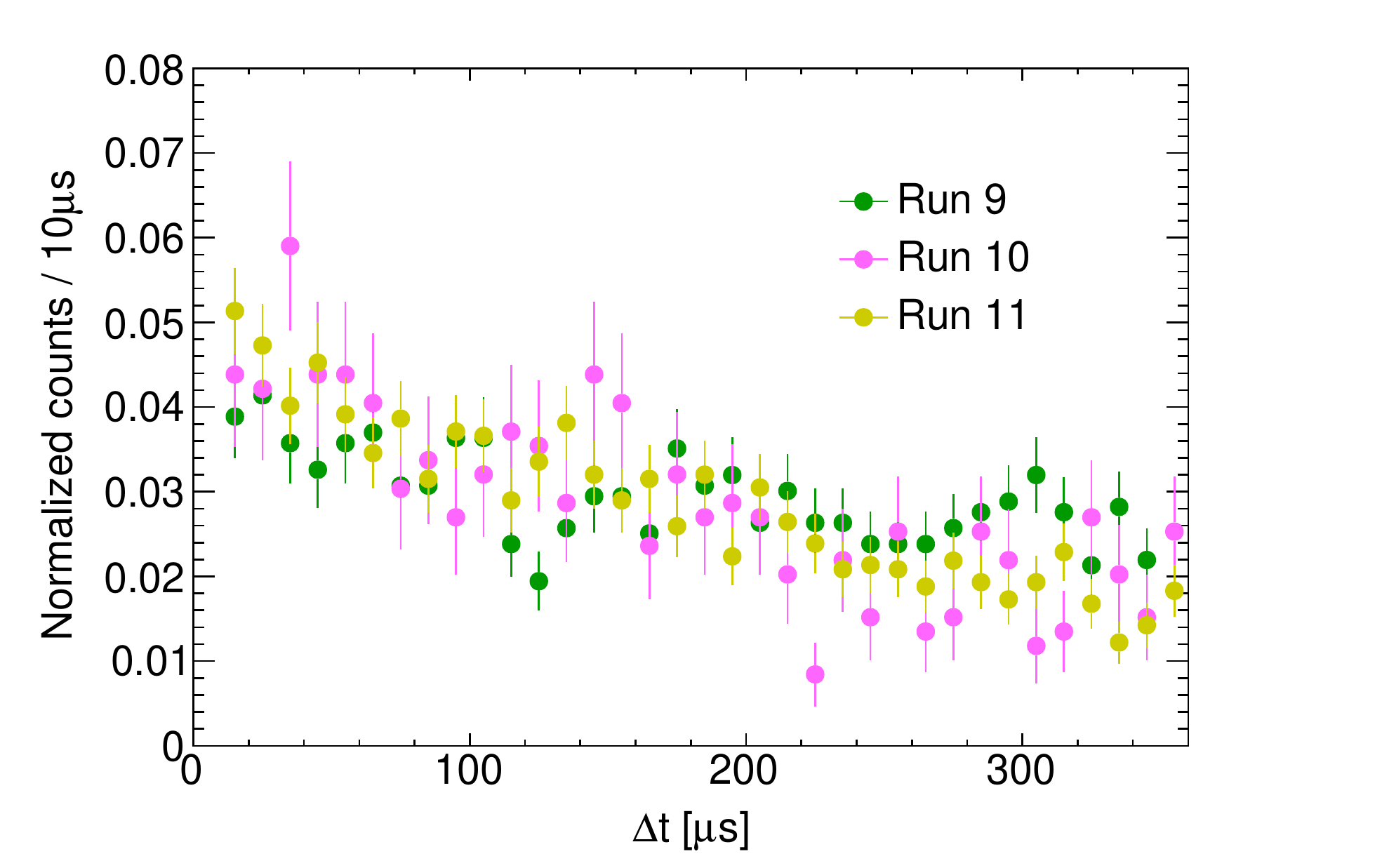}
  \caption{Distribution of the time difference $\Delta t$ between the
    isolated $S1$ and the triggered $S1$ in method 2.}
  \label{fig:dt_m2}
\end{figure}

In this method, the isolated $S1$ is defined as small $S1$ signals
before the triggered $S1$, which has no paired $S2$ within the window
of maximum drift time (see Figure~\ref{fig:s1_m2}). The triggered $S1$
should be larger than 100~PE. The time difference $\Delta t$ (see
Figure~\ref{fig:dt_m2}) between the isolated $S1$ and the triggered
$S1$ is used directly in the simulation of accidental background by
pairing the selected isolated $S1$ and $S2$ signals (see following
section), therefore we require that $\Delta t$ should be within the
window of $(10, 350)$~$\mu$s for Run 9 or $(10, 360)$~$\mu$s for Runs
10 and 11 before the triggered $S1$, respectively, by considering the
cut on the drifting time.

The rate $r_1$ of isolated $S1$ in one data taking period, each
consisting of several adjacent runs with nearly identical running
conditions, can be estimated by
\begin{equation}
  \label{eq:is1_rate_m2}
  r_{1} = \frac{n_{iS1}}{n_{tS1}}\cdot \frac{1}{\Delta t_{AB}},
\end{equation}
where $n_{iS1}$ is the number of isolated $S1$, $n_{tS1}$ is the
number of events triggered by unpaired $S1$, and $\Delta t_{AB}$ is
size of the time window, which equal to $340~\mu$s for Run 9, and
$350~\mu$s for Runs 10 and 11, respectively. The data taking periods
have similar duration.

This method was used in the first analysis of
PandaX-II~\cite{PandaX-II:2016vec}. By studying the distribution of
$\Delta t$ in Figure~\ref{fig:dt_m2}, we found that the number of
events decreased with the increasing $\Delta t$, indicating the
possible physical correlation between some selected $S1$s. This
phenomena becomes obvious in Run 11 due to the long data taking time.
The correlation may come from the $^{214}$Bi-$^{214}$Po cascade decay
in the region below the cathode (region 2 in Figure~\ref{fig:tpc}). A
half-life of $173.59\pm12.53$~$\mu$s is obtained by fitting the decay
component of the time distribution, and the value is consistent with
the half-life of $^{214}$Po (164~$\mu$s). Thus the hypothesis is
supported, and method 2 results in an over-estimated rate of isolated
$S1$. The average rate could be corrected by subtracting the
contribution from the $^{214}$Bi-$^{214}$Po events, with additional
uncertainty introduced by the correction.

\subsubsection{Method 3}
\label{sec:s1_m3}
\begin{figure}[hbt]
  \centering
  \includegraphics[width=0.8\textwidth]{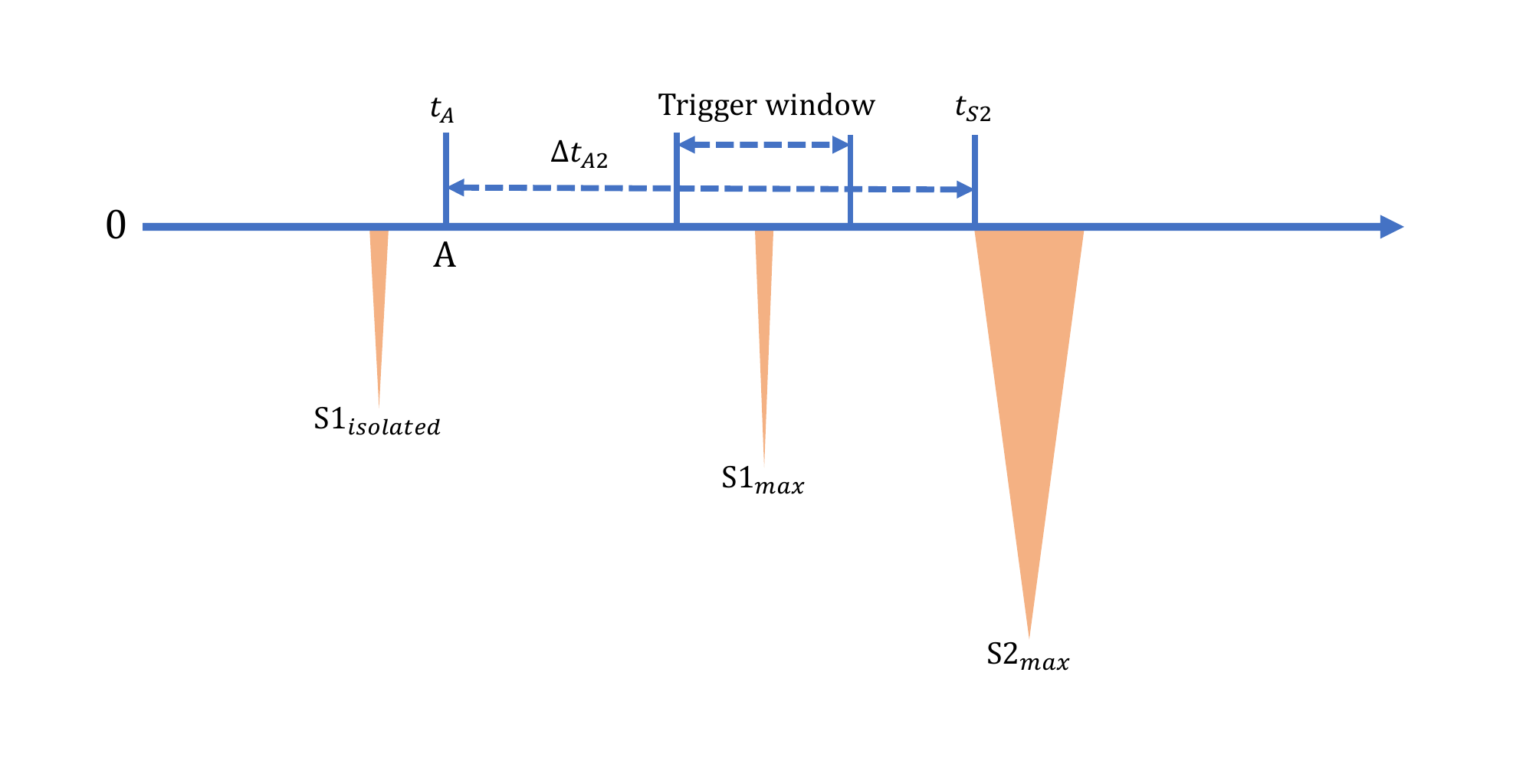}
  \caption{Schematic view on the search of isolated $S1$ in events
    triggered by $S1$ ($S1_{max}$) in Runs 10 and 11.}
  \label{fig:s1_m3}
\end{figure}

This method searches for isolated $S1$ before a good event, which is
triggered by $S1$ signal larger than 100~PE and paired with $S2$
larger than $10,000$~PE (see Figure~\ref{fig:s1_m3} for details). The
isolated $S1$ is required to be before the maximum drift time of the
$S2$ signal, i.e., 350~$\mu$s for Run 9 and 360~$\mu$s for Runs 10 and
11, to ensure no correlation between the isolated $S1$ and the $S2$ in
the good event. The cascaded decays of $^{214}$Bi-$^{214}$Po could not
enter into the data selection because two large $S2$ signals are
expected if they happen in the sensitive region.

In this method, the rate $r_1$ in a data taking period can be estimated as
\begin{equation}
  \label{eq:is1_rate_m3}
  r_1 = \frac{n_{iS1}}{ \sum{(t_{S2} - \Delta t_{A2}})},
\end{equation}
where $n_{iS1}$ is the total number of isolated $S1$. The variables of
time are defined in each of the good event, with $t_{S2}$ as the start
time of the $S2$ signal relative to the start of the event, and
$\Delta t_{A2}$ as the size of the exclusion window, which takes the
same value as the maximum drift time.

\begin{figure}[hbt]
  \centering
  \begin{subfigure}{0.45\linewidth}
    \includegraphics[width=\textwidth]{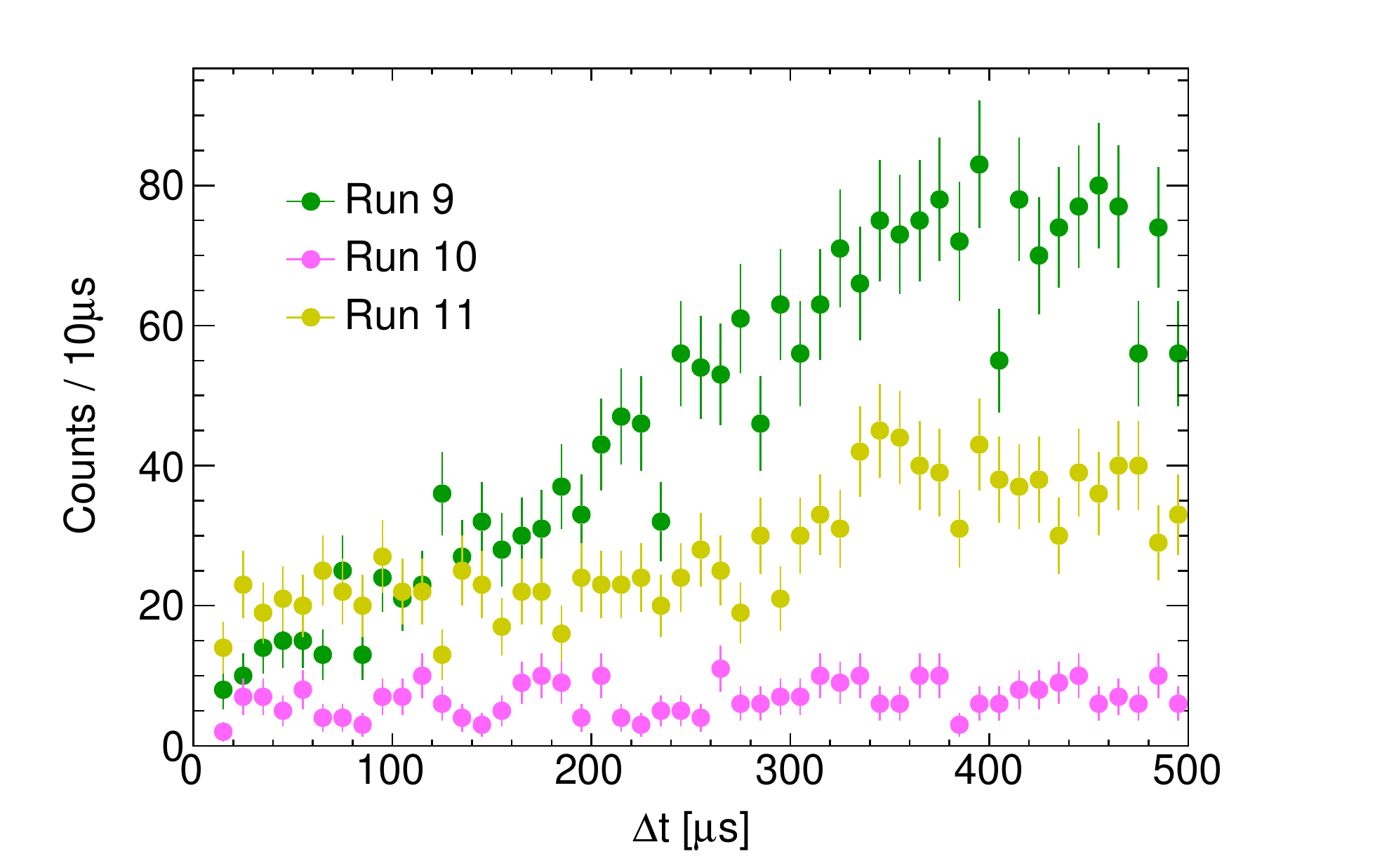}
    \caption{raw}
    \label{subfig:m3_raw}
  \end{subfigure}
  \begin{subfigure}{0.45\linewidth}
    \includegraphics[width=\textwidth]{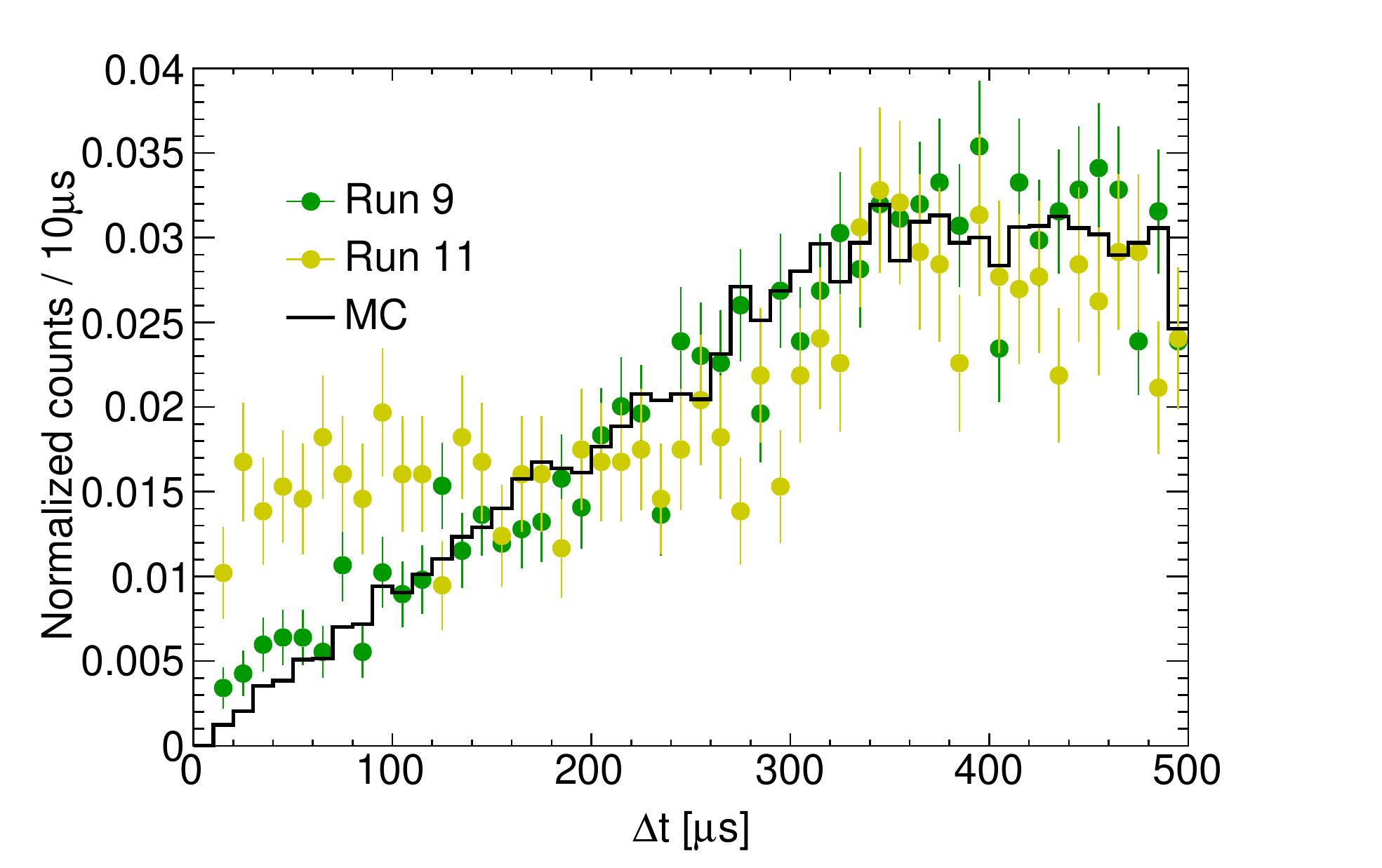}
    \caption{normalized}
    \label{subfig:m3_normalized}
  \end{subfigure}
  \caption{Distribution of the time difference $\Delta t$ between the
    isolated $S1$ and the triggered $S1$ in method
    3. \subref{subfig:m3_raw}: Raw
    distribution. \subref{subfig:m3_normalized}: The integration of
    the distribution is normalized to 1.}
  \label{fig:dt_m3}
\end{figure}

We studied the distribution of time difference $\Delta t$ between the
isolated $S1$ and the good $S1$, as shown in Figure ~\ref{fig:dt_m3}.
Considering the uniformity separation of the physical $S1$ and $S2$
signals, the requirement of the isolated $S1$ outside the maximum
drift window reduces the probability of isolated $S1$s with small
$\Delta t$ to be selected. Because the selection window
is reduced in the same time, the rate calculation is not affected. This
behavior is reproduced with a simple toy MC simulation by randomly
sampling $S2$ after the triggered $S1$ in the drift window and
randomly sampling isolated $S1$ in the whole event window, especially
for Run 9. The same MC simulation can also be used to verify the rate
calculation. Assuming the rate of isolated $S1$ is 500~Hz, the rate
calculated with method 3 is $498.9$~Hz, showing a good accuracy. For
Run 10, the behavior is not visible due to the relative low statistics
of the isolated $S1$. For Run 11, excess isolated $S1$s (~$11.6\%$)
are observed for $\Delta t < 120$~$\mu$s. They are found in the events
accumulated in the cathode region, as illustrated in
Figure~\ref{fig:m3_cathode} in Appendix~\ref{sec:appendix}. The origin
of these signals is still unknown.

\subsection{Tagging of isolated $S2$}
The estimation of the rate $r_2$ for isolated $S2$ is more
straightforward in comparison with isolated $S1$. The events triggered
by unpaired $S2$, with all the related quality cuts applied, are
selected to calculate the rate. The rate is defined as
\begin{equation}
r_{2} = \frac{n_{iS2}}{T},
\label{eq:is2_rate}
\end{equation}
where $n_{iS2}$ is the number of events satisfying the selection
criteria, and $T$ is the duration of the run.

\subsection{Properties of isolated signals}
\label{sec:properties_signals}
The estimated average rates of isolated $S1$ and $S2$ in each run are
presented in Table~\ref{tab:rate}. Run 9 has the highest rate of
isolated $S1$, which is very likely to be attributed to the higher
dark rate of PMTs operating with higher gain~\cite{Cui:2017nnn}. For
Runs 10 and 11, the $\bar{r}_1$ calculated with method 1 and method 3
are consistent with each other within uncertainty. The results of
method 3 are used in the final analysis of
PandaX-II~\cite{Wang:2020coa} and rest of this study to estimate the
rate of accidental background. The variance of the average rates of
isolated $S2$ is small.

\begin{table}[h]
  \centering
  \begin{tabular}{cccccc}
    \multirow{2}*{Run}& \multirow{2}*{Duration [days]} & \multicolumn{3}{c}{$\bar{r}_1$ [Hz]} & \multirow{2}*{$\bar{r}_2$ [Hz]} \\
    \cline{3-5} & & Method 1 & Method 2 & Method 3 \\
    \hline
    9 & 79.6  & - & $1.40\pm0.25$ & $1.53\pm0.16$ & $0.0121\pm0.0002$ \\

    10 & 77.1  & $0.46\pm0.05$ & $0.27\pm0.20$ & $0.47\pm0.02$ & $0.0130\pm0.0007$ \\
    
    11 & 244.2 & $0.77\pm0.06$ & $0.37\pm0.16$ & $0.69\pm0.06$ & $0.0121\pm0.0001$ \\\hline
    
  \end{tabular}
  \caption{Average rates of isolated $S1$ and $S2$ extracted from
    PandaX-II data. The results from method 2 have been corrected by
    subtracting the contamination from the possible
    $^{214}$Bi-$^{214}$Po cascade decay signals.}
  \label{tab:rate}
\end{table}

\begin{figure}[hbt]
  \centering
  \includegraphics[width=0.85\textwidth]{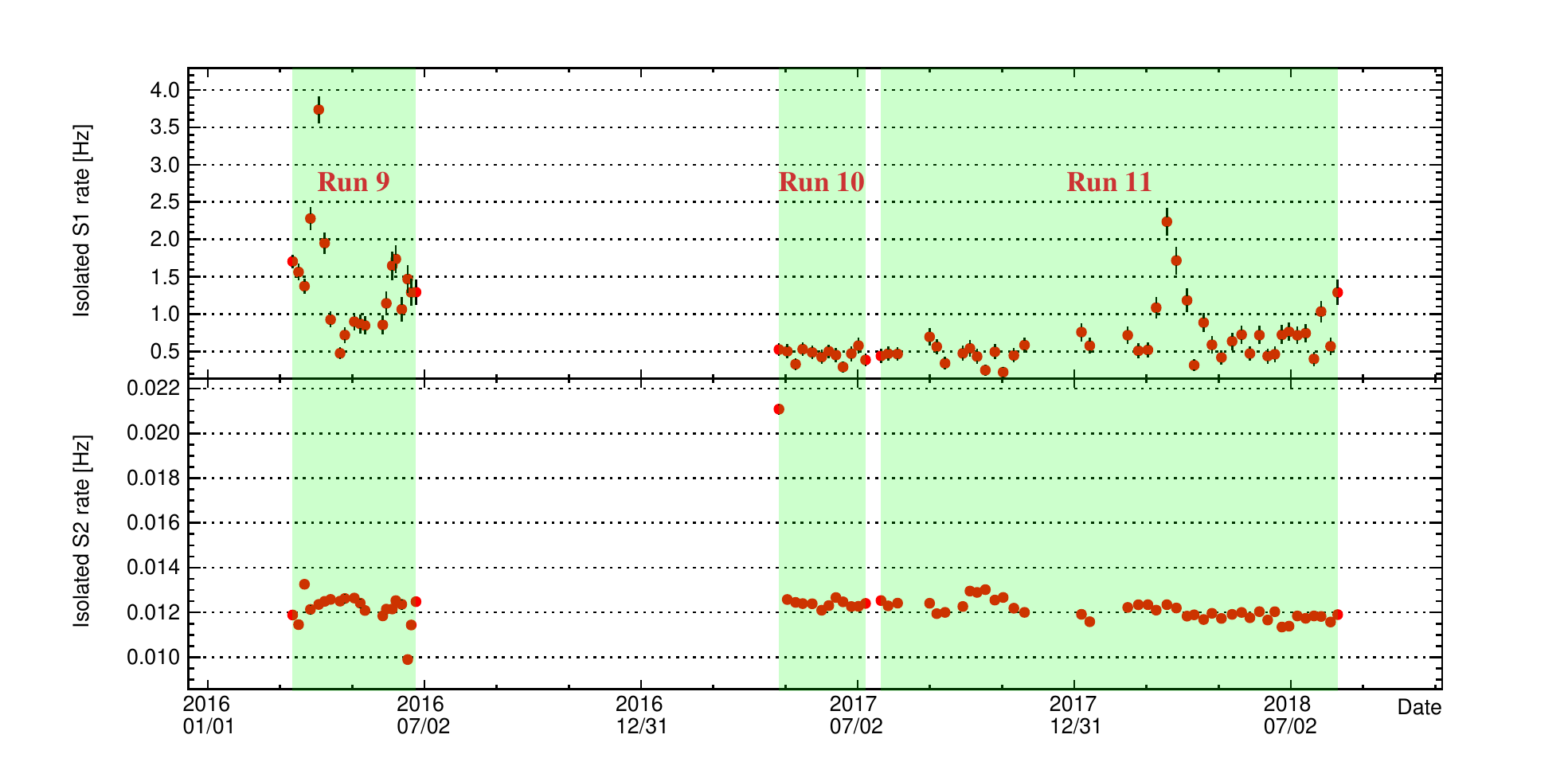}
  \caption{Evolution of rates of the isolated signals during the
    whole PandaX-II data taking period, selected by method 3.}
  \label{fig:rate_evolution}
\end{figure}

More detailed evolution of rates of the isolated signals during the
whole PandaX-II data taking period, with those of isolated $S1$
calculated by method 3, is presented in
Figure~\ref{fig:rate_evolution}. The rate of isolated $S2$ keeps
stable, while that of isolated $S1$ varies greatly. The large variance
of $r_1$ in Run 9 might come from the occasional sparking of
electrodes or PMTs. A peak rate of isolated $S1$ is observed in Run
11, which can be explained by the fact that some PMTs were unstable
during the corresponding period, as shown in
Figure~\ref{fig:hitpattern_high_rate} in Appendix~\ref{sec:appendix}. The
ordinary data quality cut cannot remove related events efficiently.

\begin{figure}[hbt]
  \centering
  \begin{subfigure}{0.45\linewidth}
    \includegraphics[width=\textwidth]{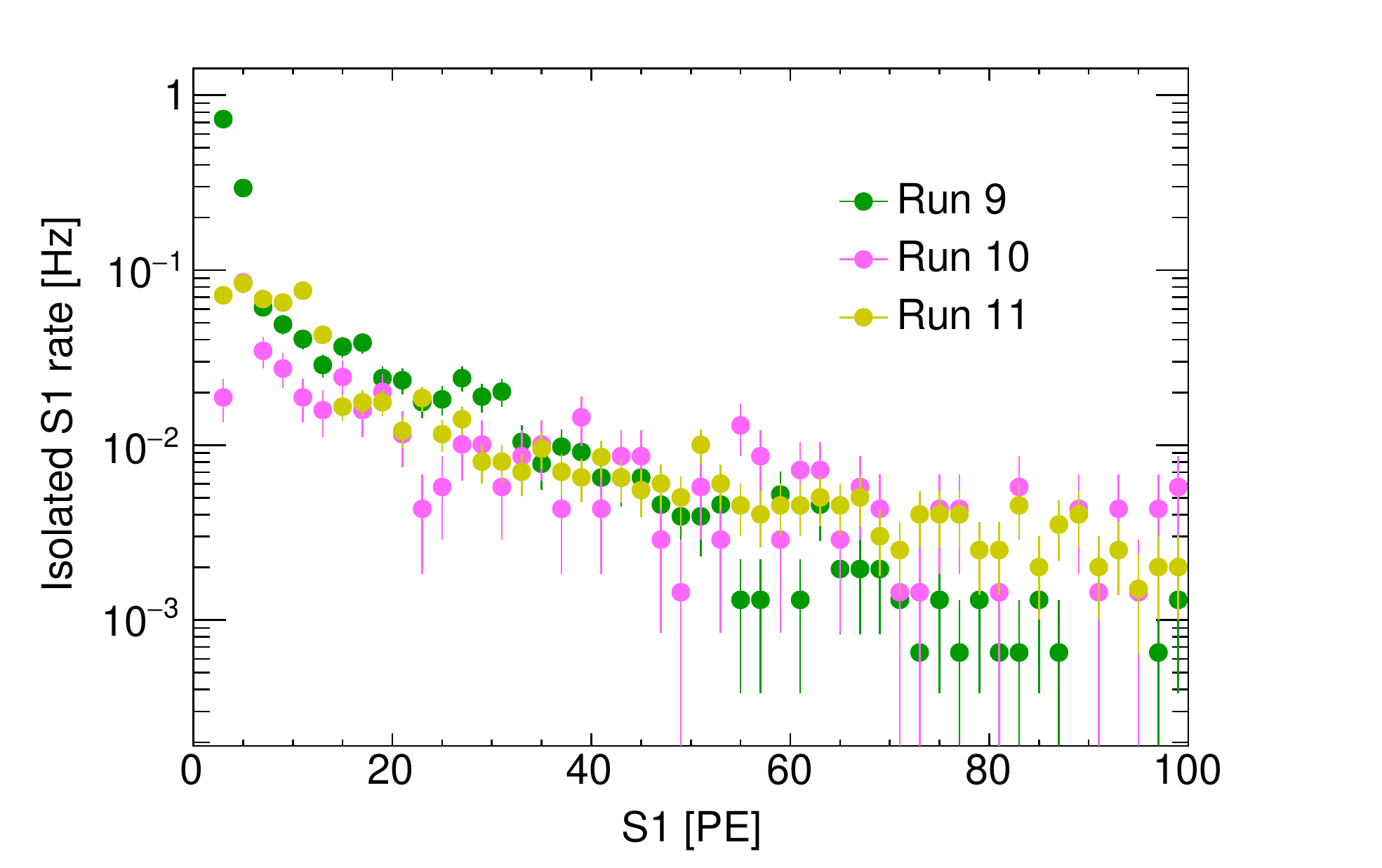}
    \caption{Isolated $S1$}
    \label{fig:s1_spec}
  \end{subfigure}
  \begin{subfigure}{0.45\linewidth}
    \includegraphics[width=\textwidth]{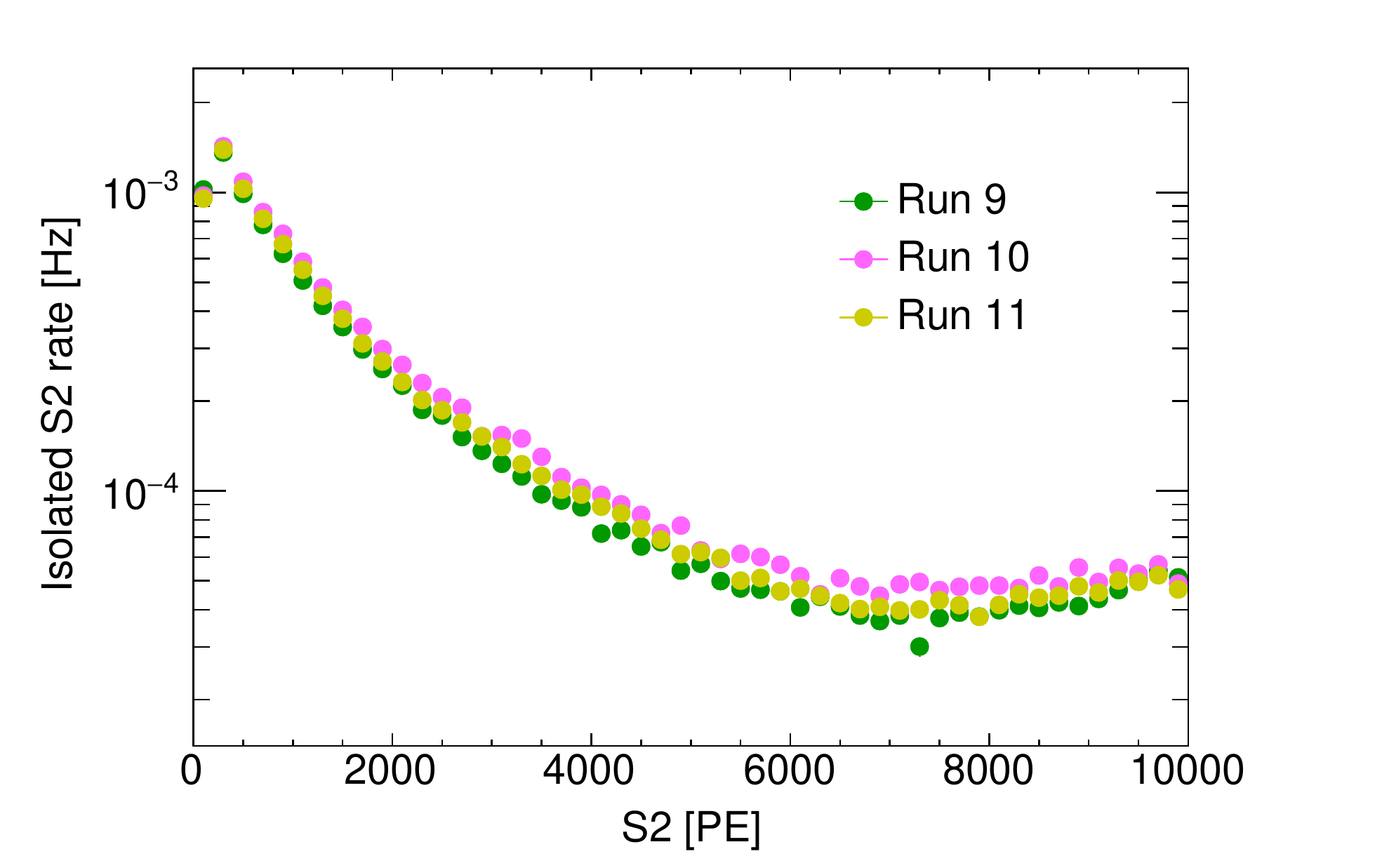}
    \caption{Isolated $S2$}
    \label{fig:s2_spec}
  \end{subfigure}
  \caption{Charge spectra of isolated signals selected by method 3.}
  \label{fig:charge_spec}
\end{figure}

The charge spectra of isolated signals selected by method 3 are shown
in Figure.\ref{fig:charge_spec}. Most of the isolated $S1$ are found
to be smaller than 10~PE. All the $S1$ spectra have similar shape when
the charge is larger than 6~PE, but a higher peak is observed below 6
PE for Run 9. This may be explained by the higher chance of accidental
coincidence of hits from dark current in this run due to the higher
operation voltage of the PMTs. A small peak in Run 11 around 10~PE is
resulted from the unstable PMTs mentioned before (see
Figure~\ref{fig:hitpattern_10pe} in Appendix~\ref{sec:appendix}). The
spectra of isolated $S2$ are consistent with each other.

\subsection{Study of the accidental background with simulation}
\label{sec:simulation_acc_bkg}
A data-driven MC simulation with the selected isolated signals is used
to study the accidental background events. For each Run, the isolated
$S1$ and $S2$ are paired randomly, with the time separation between
them sampled uniformly in the time window $\Delta t_w$ defined by the
fiducial volume cut. The horizontal position of the event is
determined by the $S2$ signal. The paired mock event is treated as an
event with raw signals. The same position-dependent charge corrections
and quality cuts for dark matter search data are applied to these
events, resulted in a cut efficiency $\epsilon$.

Then the total number $n_{\mbox{acc}}$ of accidental background events
can be calculated by
\begin{equation}
  n_{\mbox{acc}} = \bar{r}_1\cdot \bar{r}_2 \cdot \Delta t_{w} \cdot T \cdot \epsilon.
  \label{eq:acc_number}
\end{equation}
The efficiency $\epsilon$, the total number of accidental events, and
the number of events below the median line of the NR
band from calibration data~\cite{Wang:2020coa} results, are presented
in Table~\ref{tab:nacc_count}. Run 11 has the larger number of
accidental background events due to the largest duration $T$.

\begin{table}[hbt]
  \centering
  \begin{tabular}{cccc}
    \toprule
    Run & Type & $\epsilon$ & $n_{\mbox{acc}}$ \\
    \midrule
    \multirow{2}*{9}  &  total & 21.9\% & $8.15\pm0.94$ \\
      & below NR median & 3.5\% & $1.31\pm0.15$ \\
    \multirow{2}*{10}  &  total & 25.6\% & $3.16\pm0.15$ \\
     &  below NR median & 8.5\% & $1.06\pm0.05$ \\
    \multirow{2}*{11}  & total & 18.2\% & $9.87\pm0.89$ \\
       & below NR median & 5.6\% & $2.93\pm0.27$ \\
    \bottomrule
  \end{tabular}
  \caption{Number of accidental events estimated with the selected
    isolated signals using method 3.}
  \label{tab:nacc_count}
\end{table}

\begin{figure}[hbt]
    \centering
    \begin{subfigure}{0.32\textwidth}
        \includegraphics[width=\textwidth]{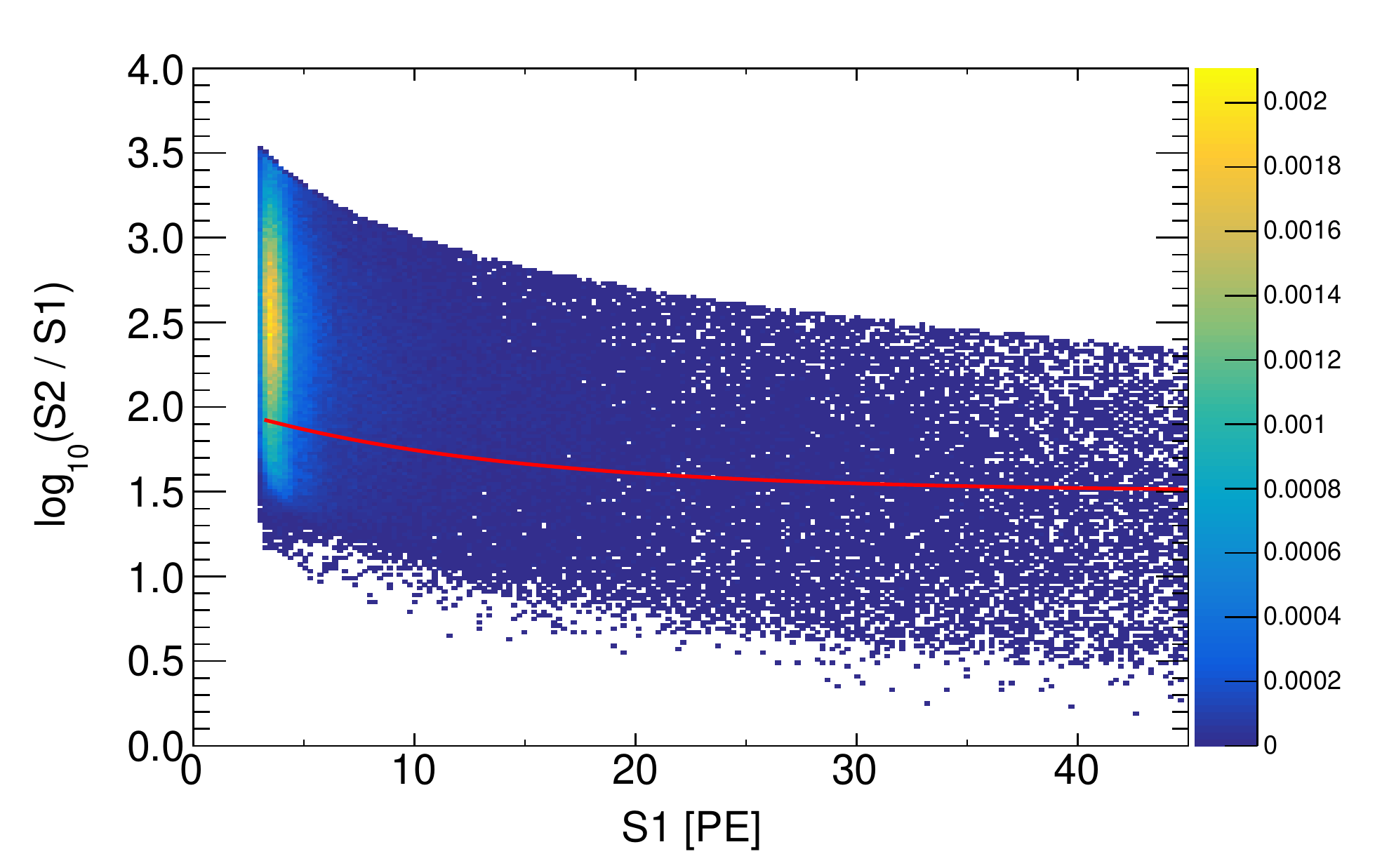}
        \caption{Run 9 Accidental}
    \end{subfigure}
    \begin{subfigure}{0.32\textwidth}
        \includegraphics[width=\textwidth]{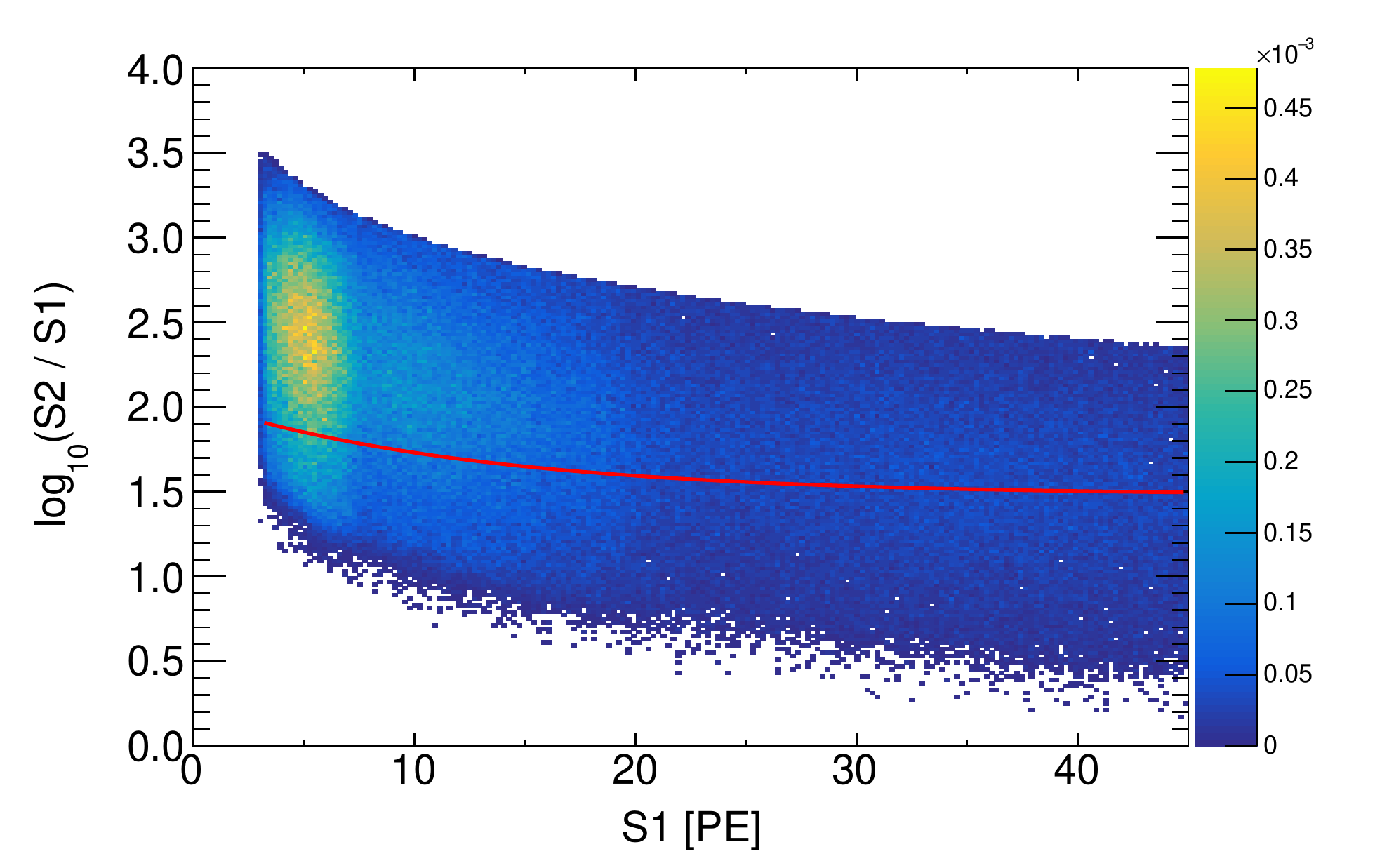}
        \caption{Run 10 Accidental}
    \end{subfigure}
        \begin{subfigure}{0.32\textwidth}
        \includegraphics[width=\textwidth]{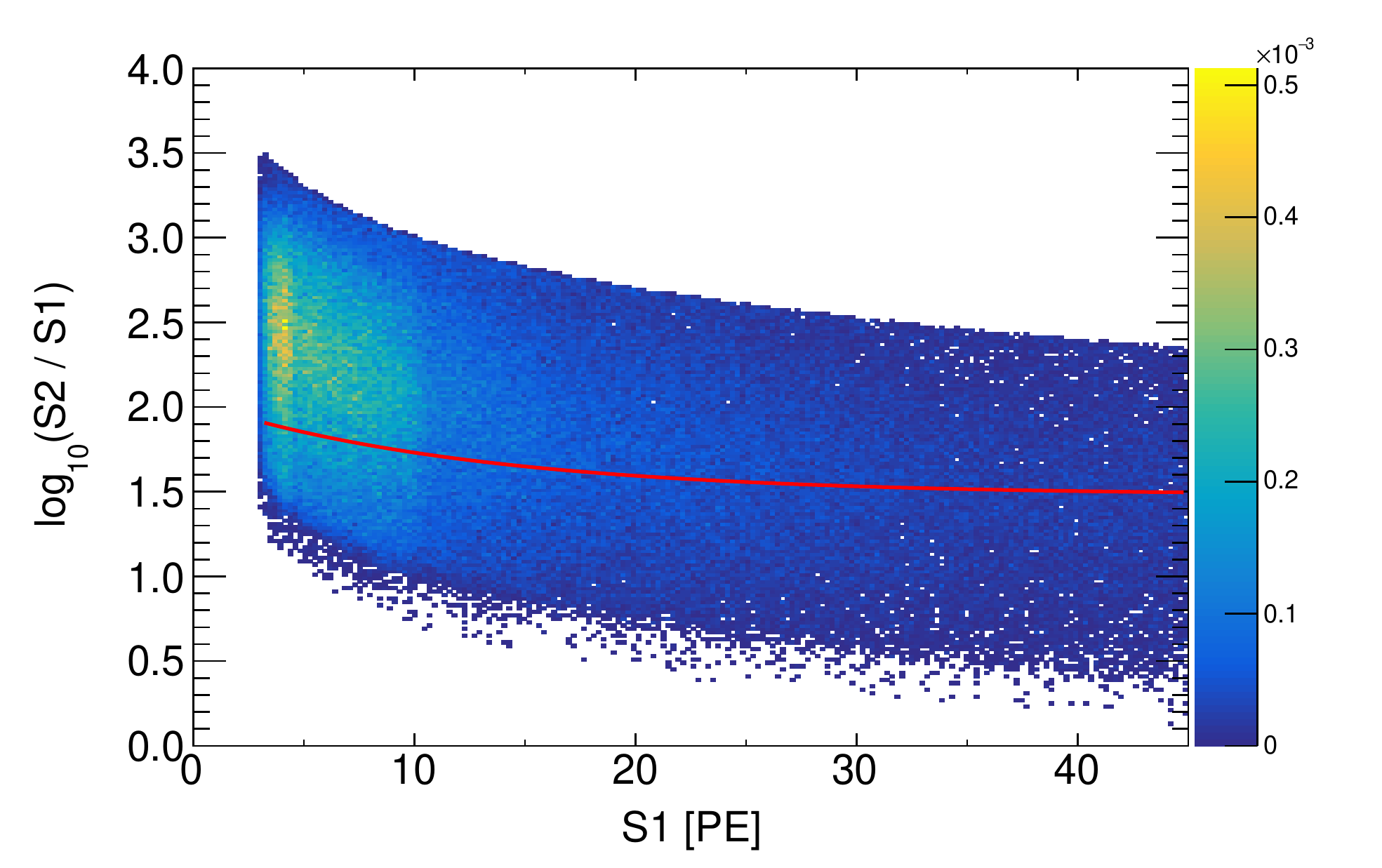}
        \caption{Run 11 Accidental}
    \end{subfigure}
    \begin{subfigure}{0.32\textwidth}
        \includegraphics[width=\textwidth]{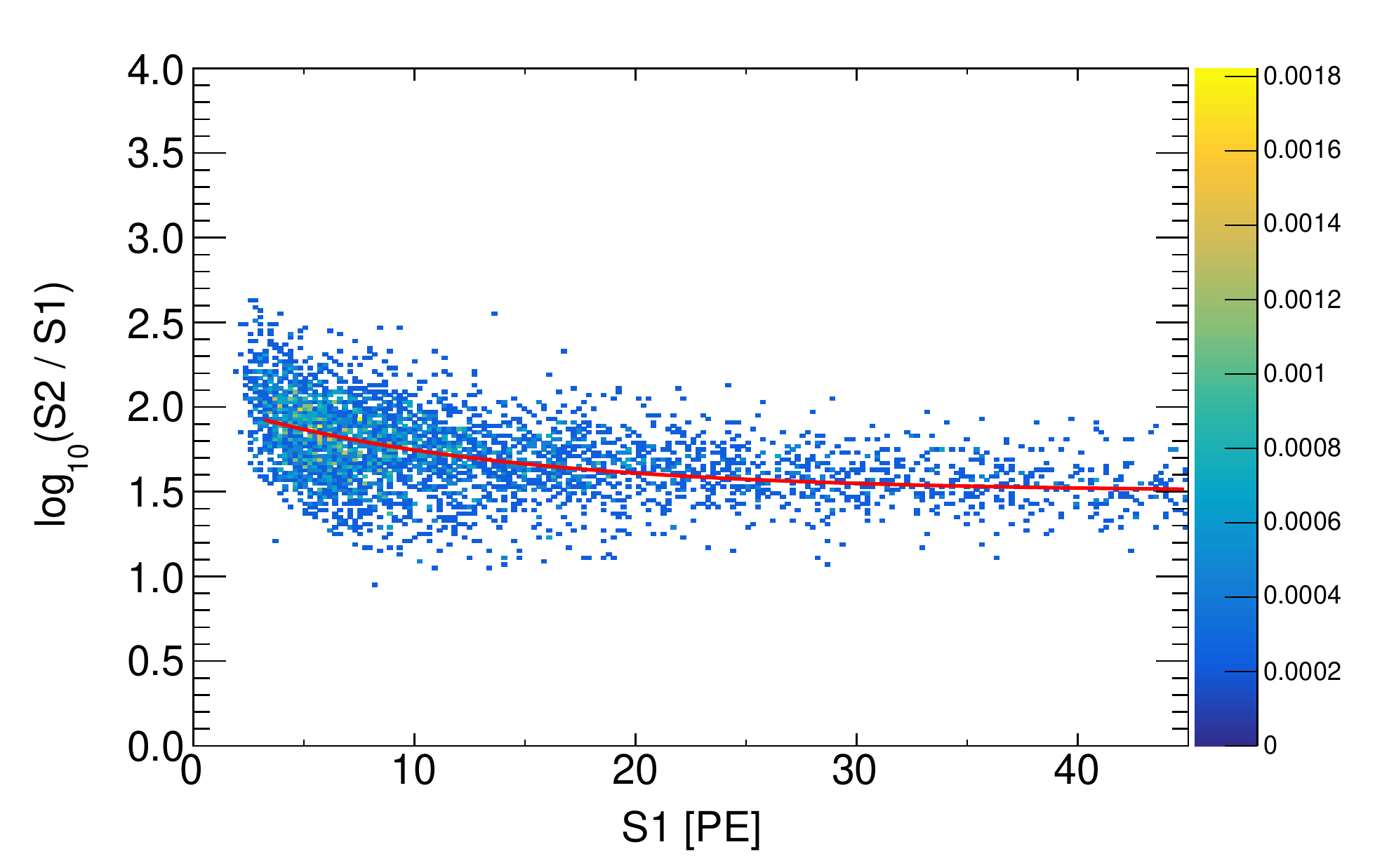}
        \caption{Run 9 NR}
    \end{subfigure}
    \begin{subfigure}{0.32\textwidth}
        \includegraphics[width=\textwidth]{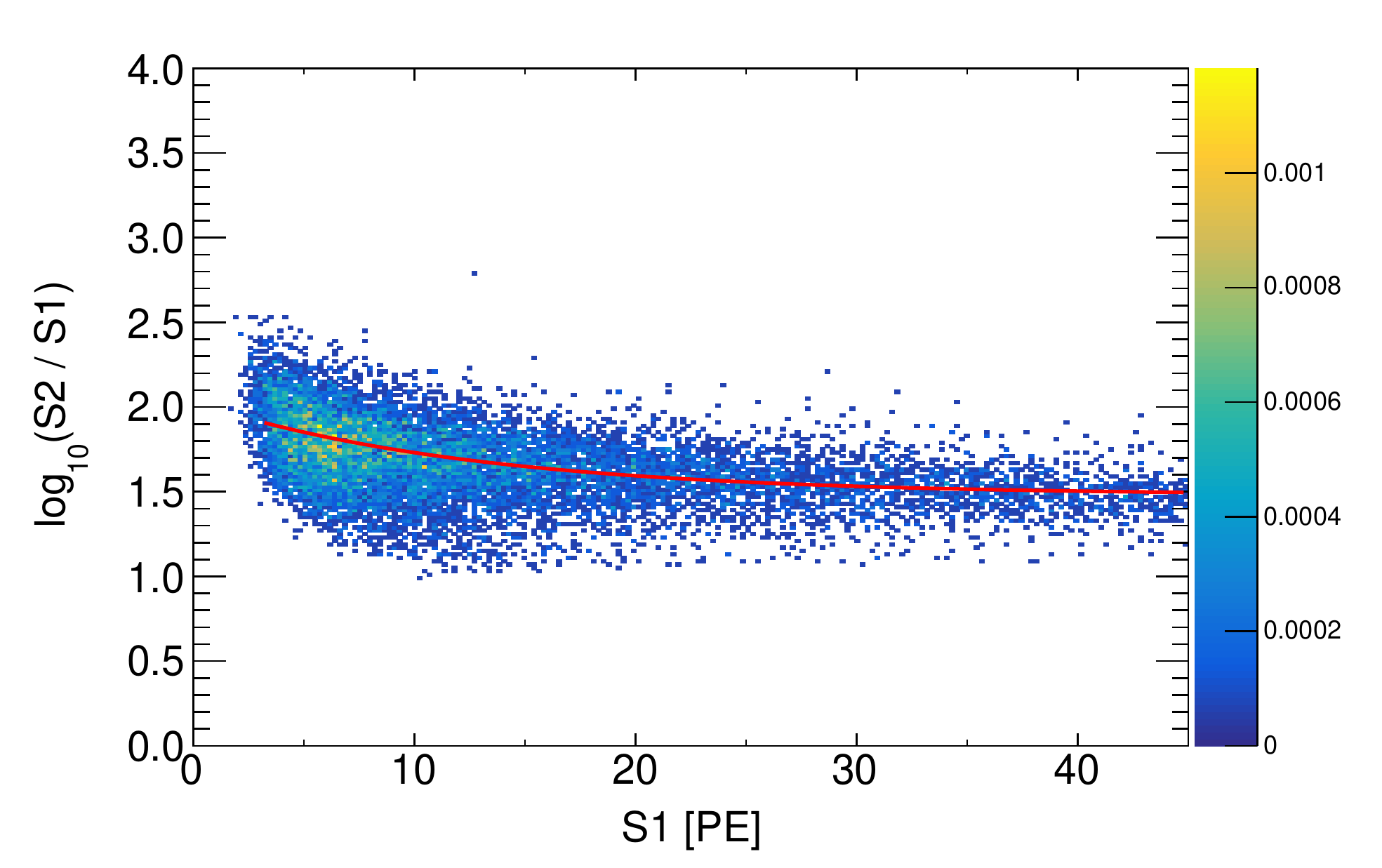}
        \caption{Run 10 NR}
    \end{subfigure}
        \begin{subfigure}{0.32\textwidth}
        \includegraphics[width=\textwidth]{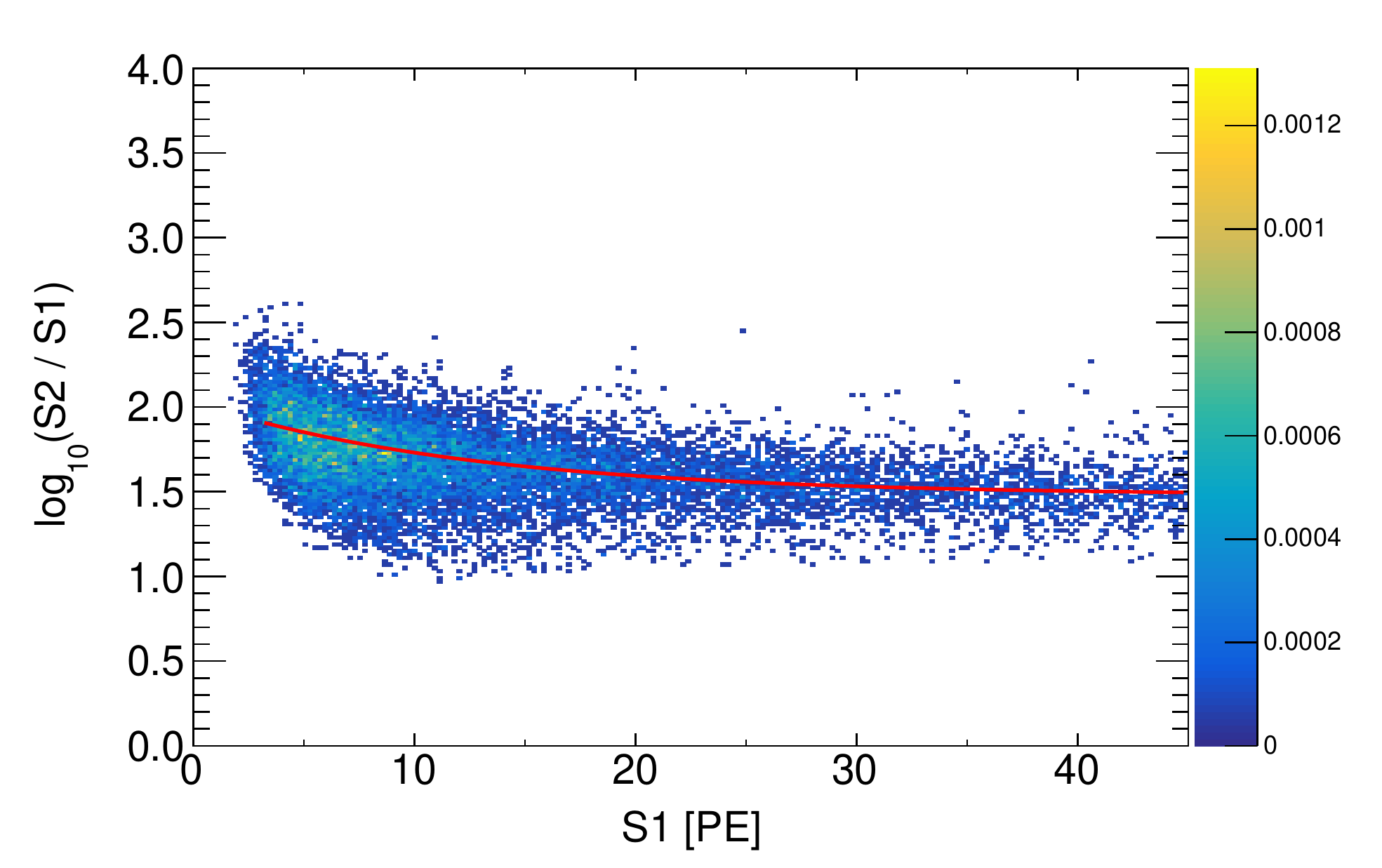}
        \caption{Run 11 NR}
    \end{subfigure}
    \caption{Distribution of $\log_{10}(S2/S1)$ vs. $S1$ for the
      simulated accidental background and NR calibration data. The
      red curves are the corresponding NR median for each Run.}
    \label{fig:charge_2d}
\end{figure}

The distributions of $\log_{10}(S2/S1)$ vs. $S1$ for the simulated
accidental background events after all the quality cuts within the
dark matter search window~\cite{Wang:2020coa} are given in
Figure~\ref{fig:charge_2d}, together with those of the NR calibration
data. Most of the accidental events have a relative small $S1$ charge
and are above the NR median.  Considering the low statistics of the
most critical ER backgrounds below the NR median, the non-negligible
accidental background in this region will reduce the discovery power
for WIMPs.  Suppressing these background could improve the sensitivity
of the detector for WIMP search.

\section{Suppression of accidental background with BDT}
\label{sec:bdt}
The accidental events are composed with isolated $S1$ and $S2$. Since
there are no physical correlation between them, we would expect a
method to distinguish them from the physical events by considering the
joint distributions of the properties of these signals. Because all
the selected accidental events have passed the quality cuts, it is
hard to tell the difference between any single property of a signal
from the accidental events and the physical events. A multi-variant
analysis could be used.  The algorithm of BDT, as one of the most
successful multi-variant analysis method used in particle
physics~\cite{Roe:2004na}, was firstly used to suppress the accidental
background in the first analysis results of
PandaX-II~\cite{PandaX-II:2016vec}. The real signal of the
WIMP-nucleon scattering is NR, thus the single scattering events
from NR calibration runs (AmBe) should be used as input signals in the
machine learning, with randomly paired events as backgrounds. Given
the fact that the ER events dominate the region above the NR median in
the dark matter search data and the relative low estimated number of
accidental events in the region, we only consider to distinguish the
accidental background from the physical NR events below the NR median.

\subsection{Variables}
\label{sec:vars}
\begin{figure}[hbt]
  \centering
  \includegraphics[width=\textwidth]{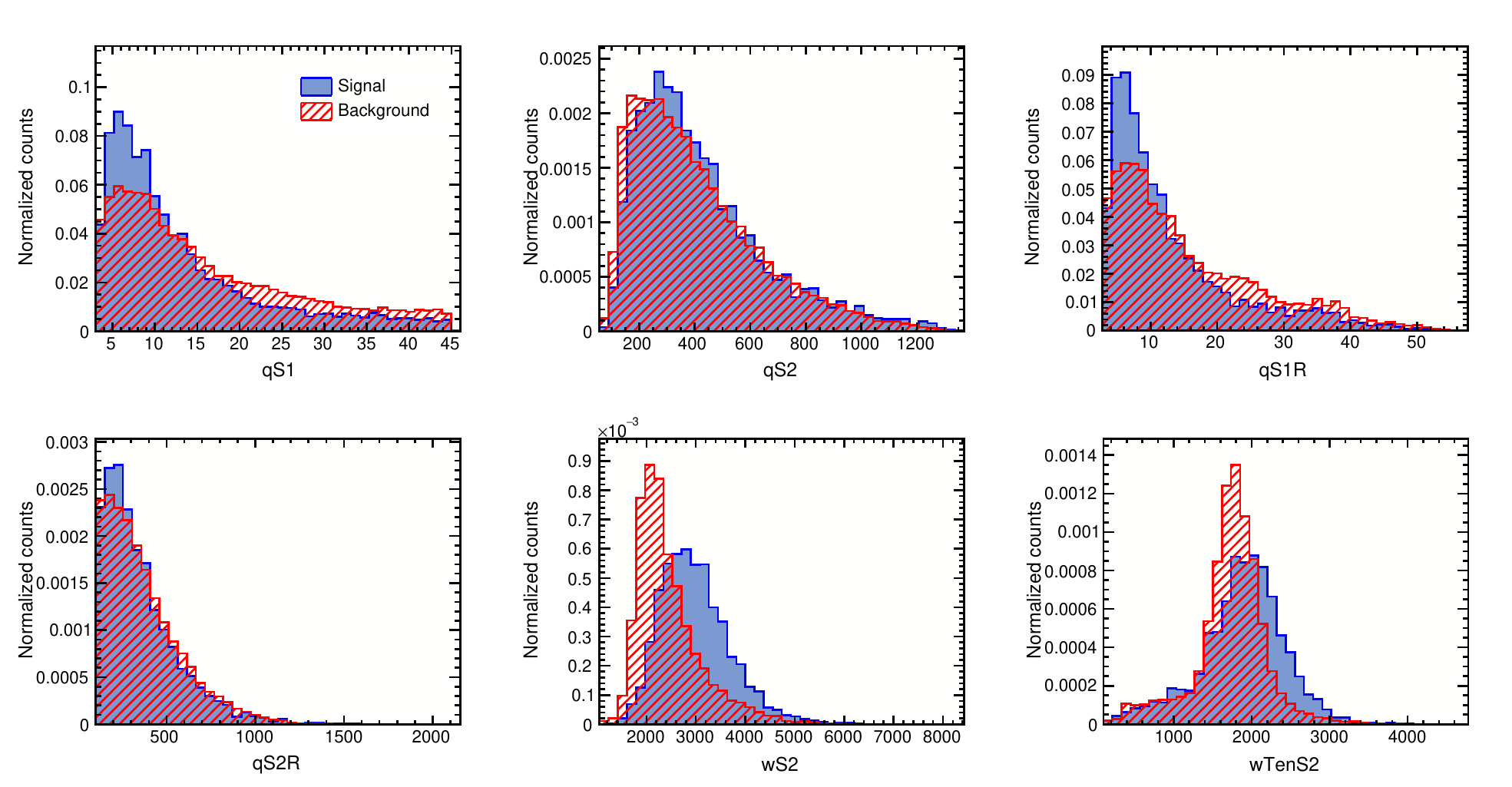}
  \includegraphics[width=\textwidth]{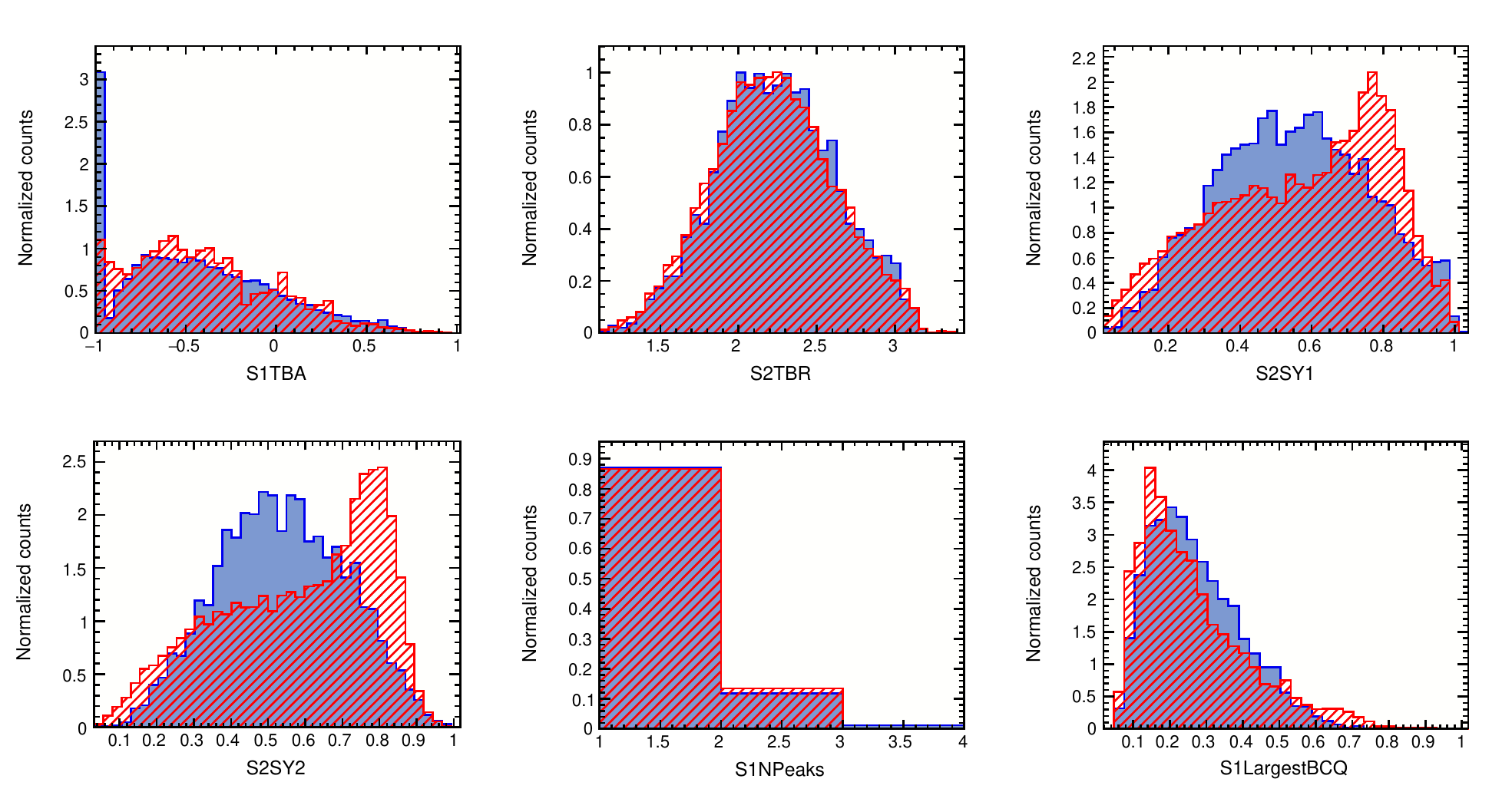}  
  \caption{Distribution of the selected variables from the NR
    calibration data (signal) and the simulated accidental events
    (background) in Run 11. Only the events below the NR median are
    selected.}
  \label{fig:var_dis}
\end{figure}

The TMVA (Toolkit for Multivariate Data Analysis) package in ROOT is
used to perform the BDT machine learning~\cite{Hocker:2007ht}.  A set
of signal properties are exploited to search for the difference
between the accidental events and the physical NR events, including
\begin{itemize}
\item corrected charge of $S1$ (qS1);
\item corrected charge of $S2$ (qS2);
\item raw charge of $S1$ (qS1R);
\item raw charge of $S2$ (qS2R);
\item width of $S2$ (wS2);
\item full width at tenth maximum of $S2$ (wTenS2);
\item asymmetry between the top charge and the bottom charge for $S1$ (S1TBA);
\item ratio of the top charge to the bottom charge for $S2$ (S2TBR);
\item the ratio of the pre-max-height charge to the total charge of an $S2$
  signal (S2SY1 in the directly summed over waveform, S2SY2 in the
  smoothed waveform);
\item number of local maximums (peaks) of $S1$ (S1NPeaks);
\item ratio of the largest charge collected by the bottom PMT of $S1$ to total charge of S1 (S1LargestBCQ)
\end{itemize}
Distributions of the these variables for the events below the NR
median can be found in Figure~\ref{fig:var_dis}, and their
correlations are presented in Figure~\ref{fig:cor_mat}.

\subsection{BDT results}
\label{sec:bdt_res}
\begin{figure}[hbt]
  \centering
  \includegraphics[width=0.65\textwidth]{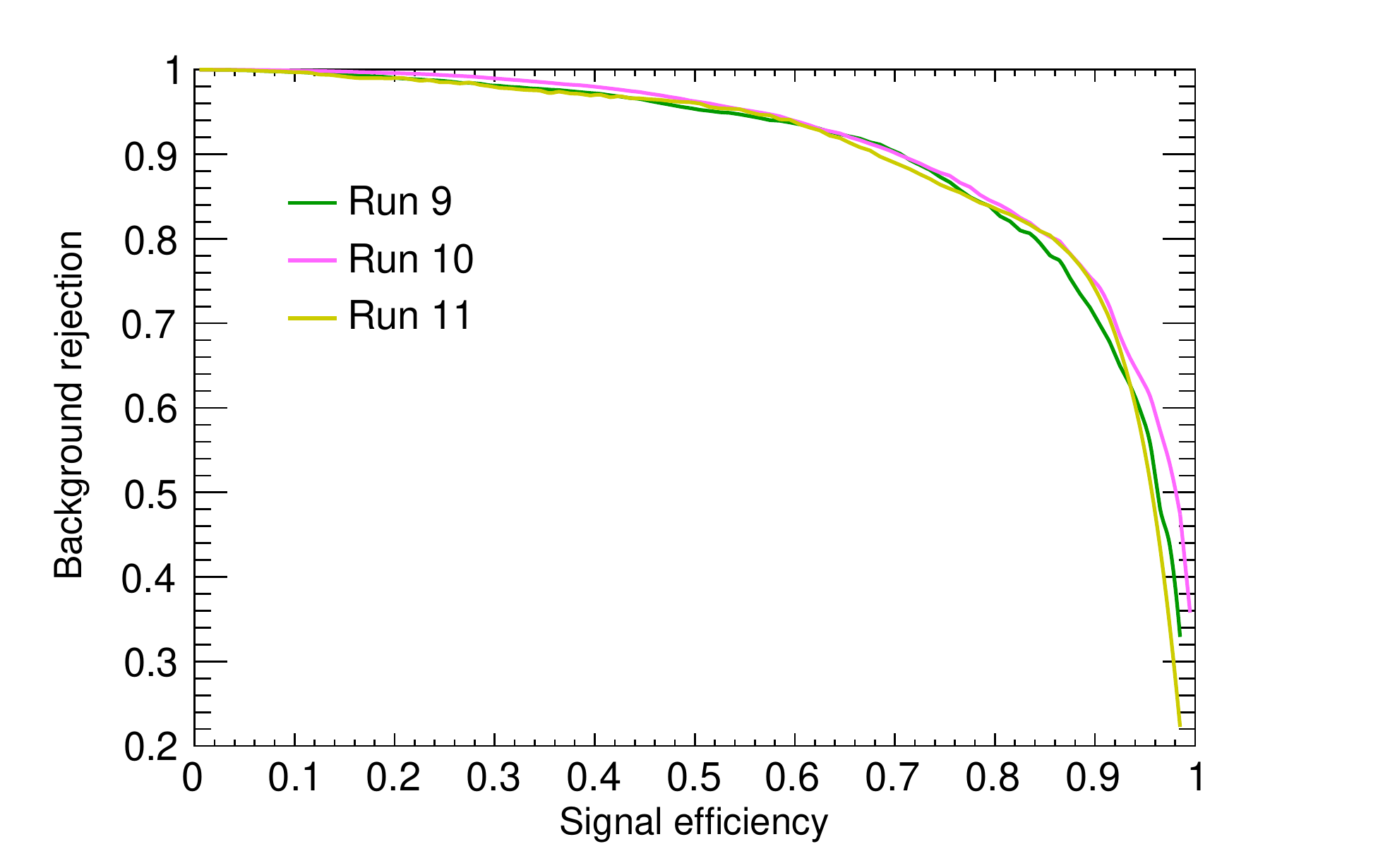}
  \caption{The evolution of the background rejection efficiency with
    the signal efficiency at different BDT cuts for different runs. The
    initial numbers of background and signal events are assumed to
    be identical.}
  \label{fig:roc}
\end{figure}

\begin{figure}[hbt]
    \centering
    \begin{subfigure}{0.32\textwidth}
        \includegraphics[width=\textwidth]{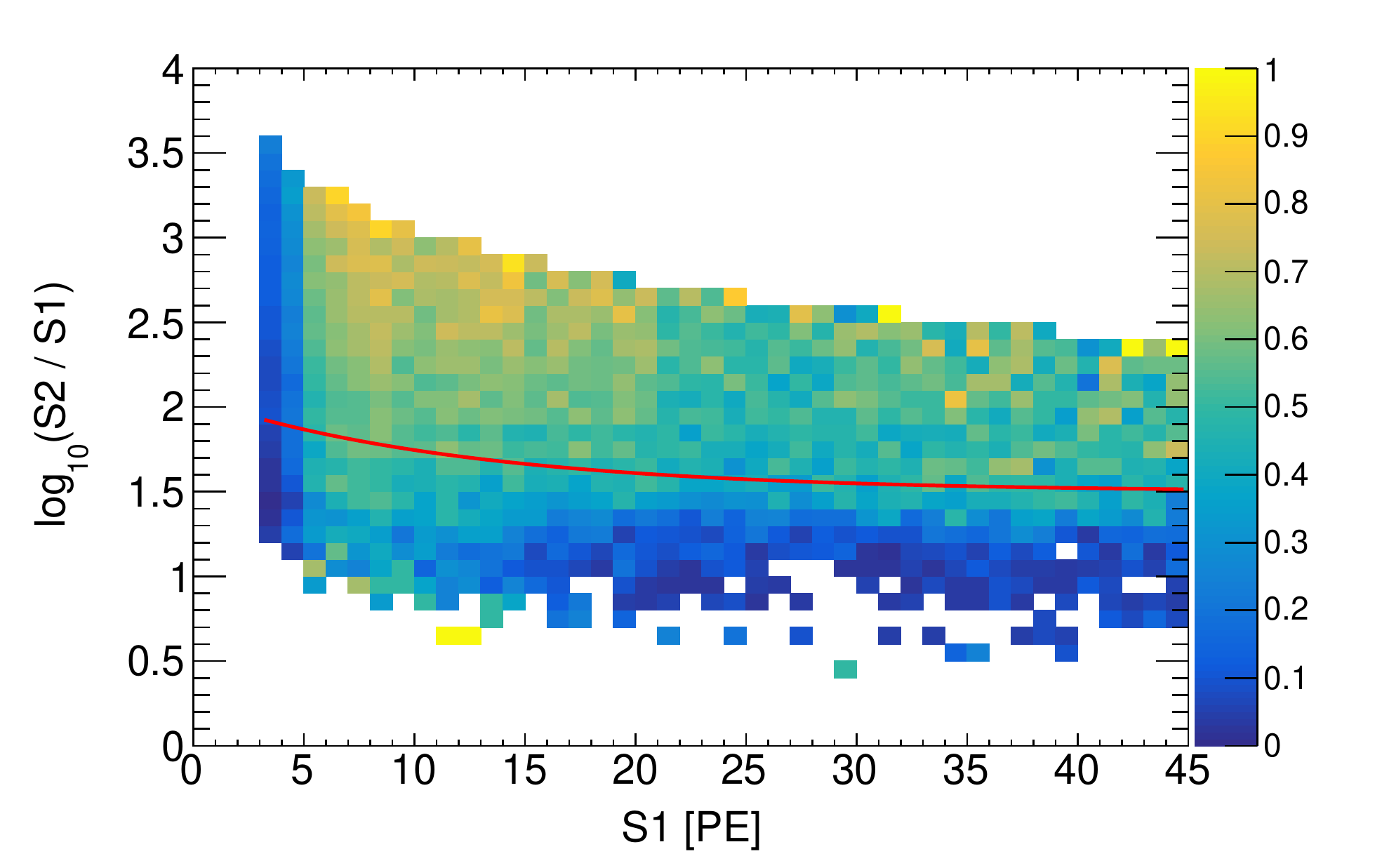}
        \caption{Run 9}
    \end{subfigure}
    \begin{subfigure}{0.32\textwidth}
        \includegraphics[width=\textwidth]{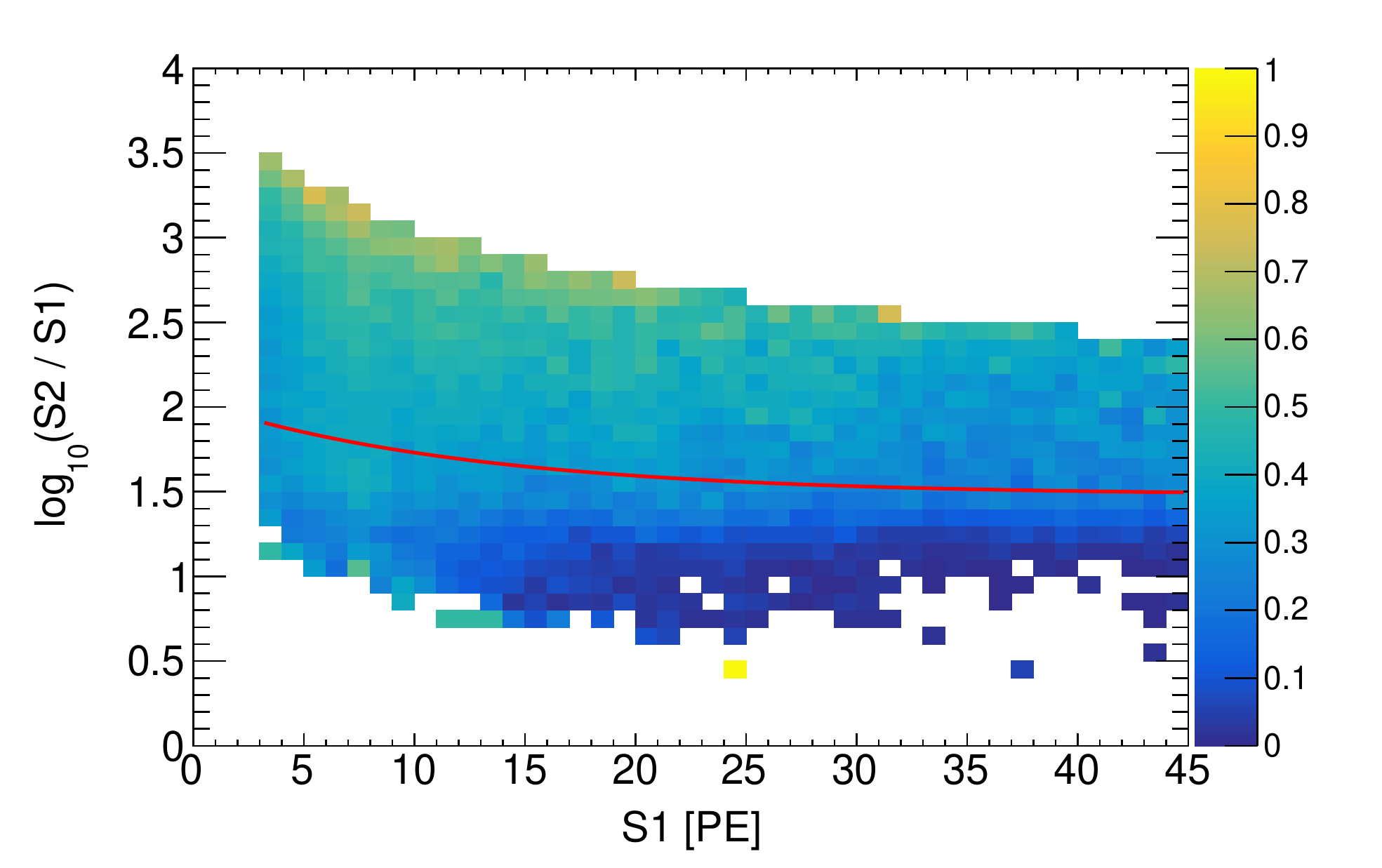}
        \caption{Run 10}
    \end{subfigure}
        \begin{subfigure}{0.32\textwidth}
        \includegraphics[width=\textwidth]{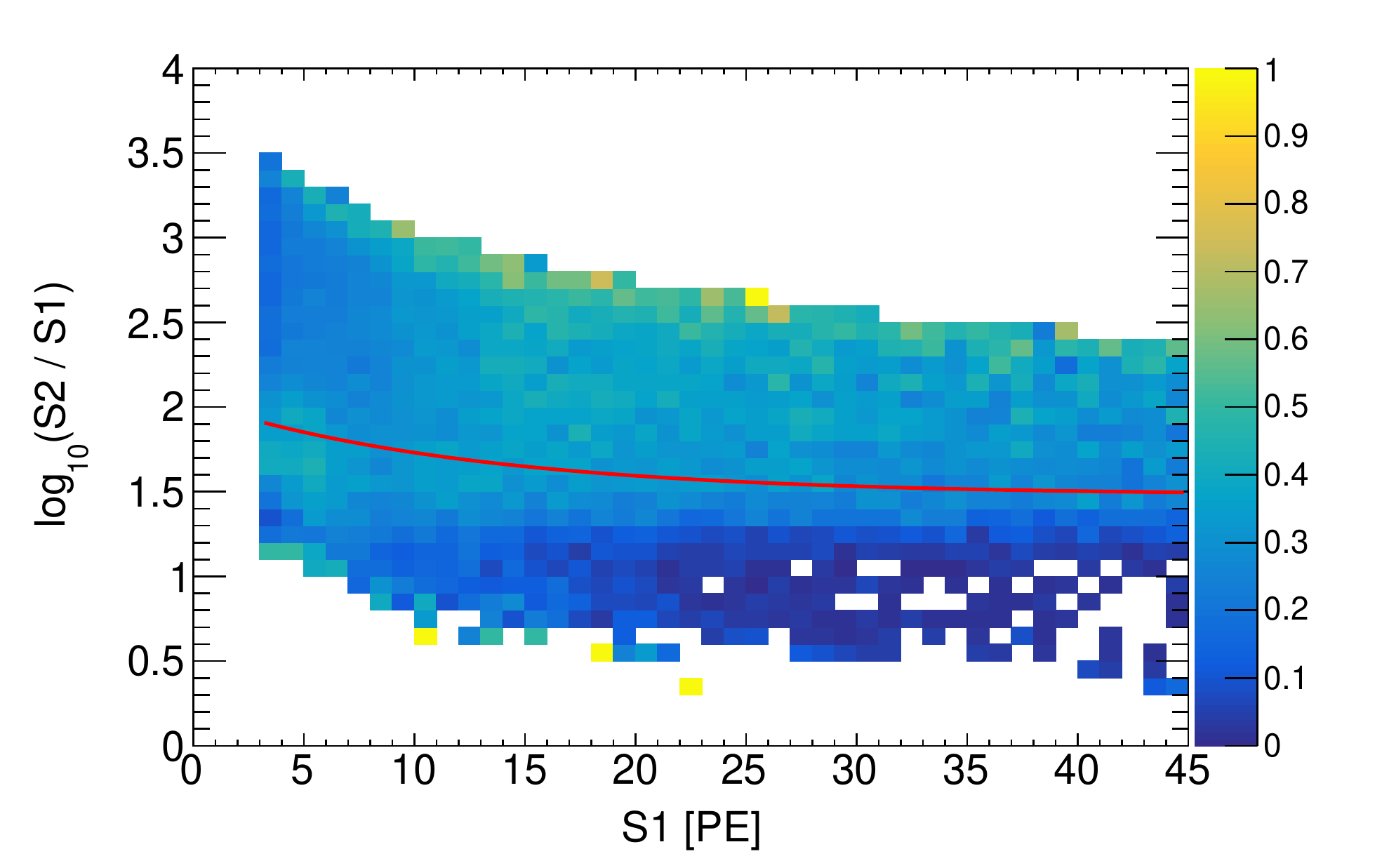}
        \caption{Run 11}
    \end{subfigure}
    \caption{
      BDT cut efficiency map on the $\log_{10}(S2/S1)$ vs. $S1$
      distribution for the simulated accidental background.}
    \label{fig:eff_2d_post_bdt}
\end{figure}

We constructed the adaptive BDT using the default parameters provided
by the official ROOT TMVA classification example, except the parameter
of \texttt{NTrees} (number of trees).  We trained the data for the
three runs independently, each with a predefined set of
\texttt{NTrees}. After the training, the resulted BDT response
distributions of the training and test data samples are superimposed
and the Kolmogorov-Smirnov (K-S) test is performed to check for
overtraining (see Fig.~\ref{fig:overtraining_check} for details). We
choose \texttt{NTrees}$\,=90$ for further study. With the trained BDT,
the ``likelihood'' estimators can be calculated for an input event to
be classified. The best cut criteria for the estimator is obtained
with the test data set by maximizing the significance $S$,
\begin{equation}
S = \frac{\epsilon_s n_s}{\sqrt{\epsilon_s n_s + \epsilon_b n_b}},
\label{eq:significance}
\end{equation}
where $n_s$ and $n_b$ are the number of signal and background events,
respectively, $\epsilon_s$ and $\epsilon_b$ are the efficiencies for
signal and background events at a given estimator value, respectively.
In this study, the expected signal events below the NR median have
high probability to be the neutron background or the WIMP events, they
are estimated at the same level as the accidental
background~\cite{Wang:2020coa}. Therefore, the identical numbers of
$n_s$ and $n_b$ are used to calculate the significance.  The evolution
of the background rejection efficiency with the signal efficiency at
different BDT cut values is shown in Figure~\ref{fig:roc}. The results
at the maximum significance $S$ are presented in
Table~\ref{tab:efficiency}. The BDT algorithm is capable to remove
$70\%$ of the accidental background events, while keeping about $90\%$
of the single scattering NR events below the NR median curve in all of
the three runs. The distributions of the BDT cut efficiency on the
$\log_{10}(S2/S1)$ vs. $S1$ plane for the simulated accidental
background are given in Figure~\ref{fig:eff_2d_post_bdt}.

\begin{table}[hbt]
  \centering
  \begin{tabular}{cccc}
    \toprule
    Run & $S$ &$\epsilon_s$& $1-\epsilon_b$ \\
    \midrule
    9 & 25.9 & 90.4\% & 70.2\% \\
    10 & 26.5 & 91.1\% & 74.6\% \\
    11 & 26.2 & 90.7\% & 73.7\% \\
    \bottomrule
  \end{tabular}
  \caption{The significance $S$, signal efficiencies $\epsilon_s$ and
    background rejection efficiencies $1-\epsilon_b$ at the best cut
    value of the estimator for events below the NR median lines, assuming $n_s=n_b$.}
  \label{tab:efficiency}
\end{table}
The contribution of each input variable to the discrimination power is
extracted by the BDT training. The variables of wS2, S2SY2, S1TBA are
found to be the most critical to the recognition of accidental
backgrounds. By checking the distributions of these variables,
isolated $S2$ signals are found to have smaller width (wS2) and more
asymmetrical shape (S2SY2) in comparison with those in normal events,
indicating that most of these signals are generated near the grid
wires~\cite{LUX:2020yym}. The peak in the S1TBA distribution of
physical events at the value of -1 suggests a large fraction of the
physical $S1$ signals have no hits on the top PMT array. Given the
fact that physical $S1$s are produced inside the liquid xenon, small
signals have smaller chance to be detected by the top PMTs due to the
total reflection on the surface between the liquid and gas xenon. But
some of the non-physical $S1$s are from the coincidence of dark noises
on the top PMTs, resulted in a S1TBA larger than -1. The distribution
of S1TBA could be used to estimate the fraction of isolated $S1$s from
the coincidence of dark noise from top PMTs. This phenomenon helps to
distinguish the non-physical small $S1$ signals from the real ones.

\subsection{Overall results}
\label{sec:overall_res}

\begin{figure}[hbt]
    \centering
    \begin{subfigure}{0.32\textwidth}
        \includegraphics[width=\textwidth]{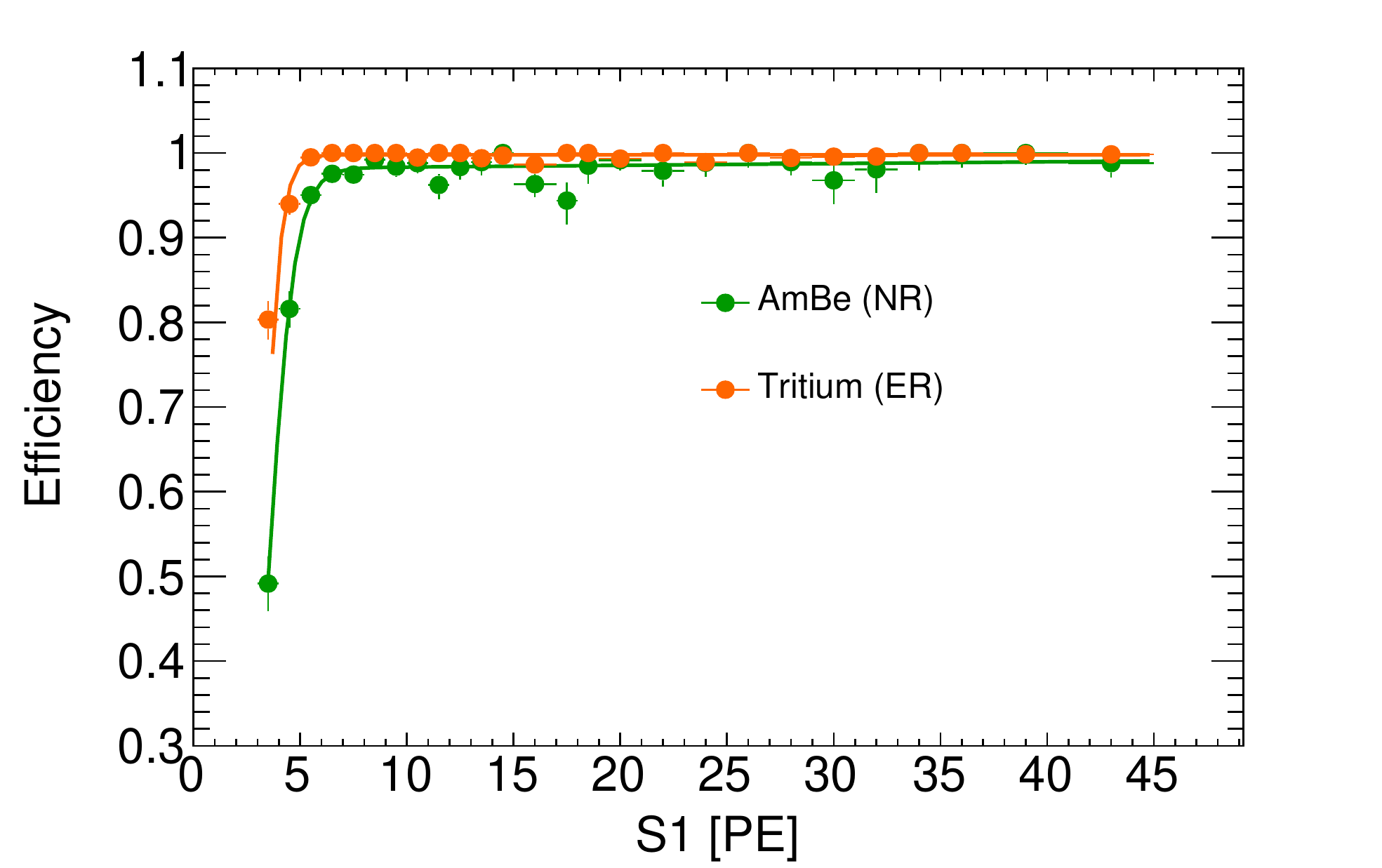}
        \caption{Run 9}
    \end{subfigure}
    \begin{subfigure}{0.32\textwidth}
        \includegraphics[width=\textwidth]{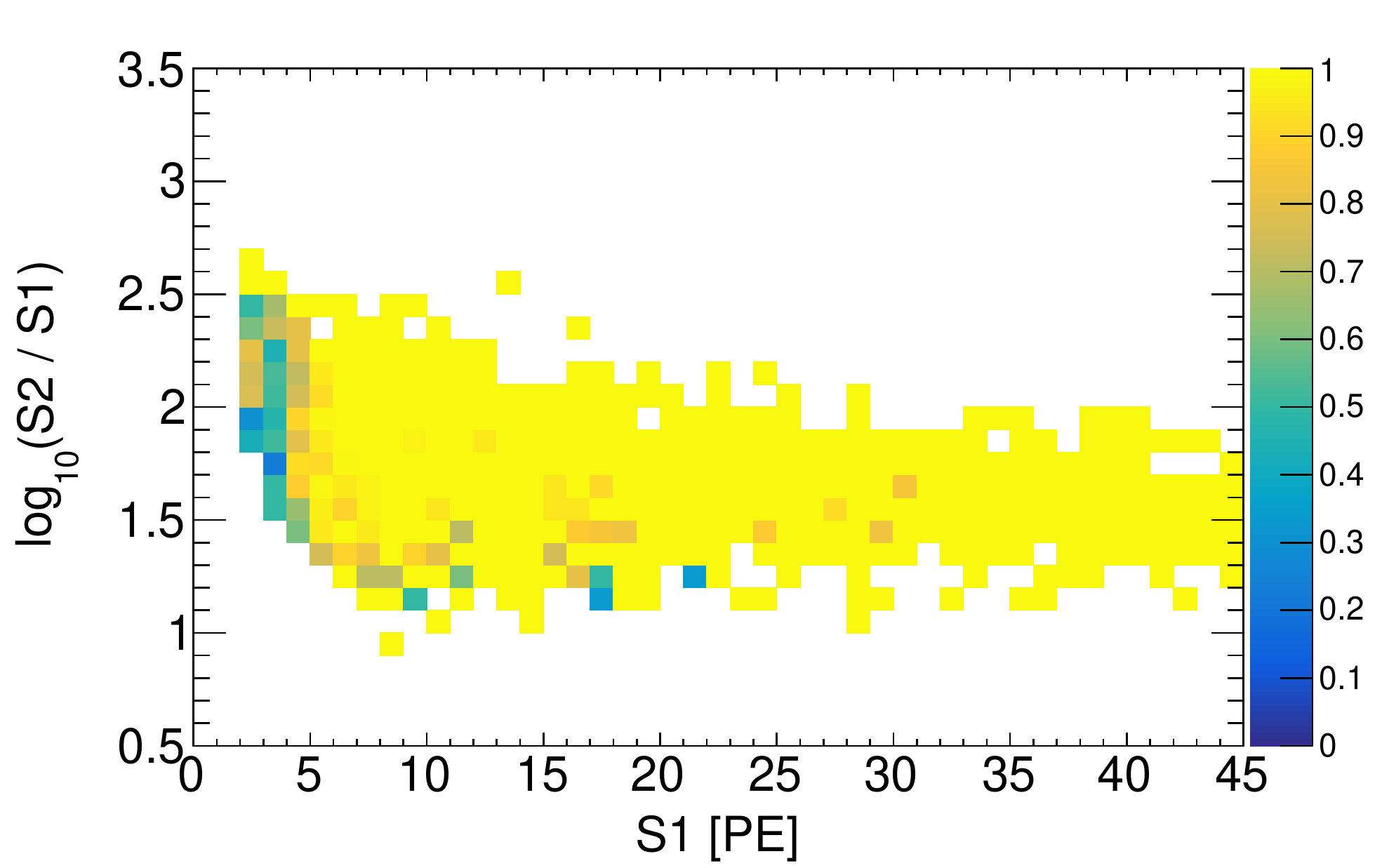}
        \caption{Run 9 (NR)}
    \end{subfigure}
    \begin{subfigure}{0.32\textwidth}
        \includegraphics[width=\textwidth]{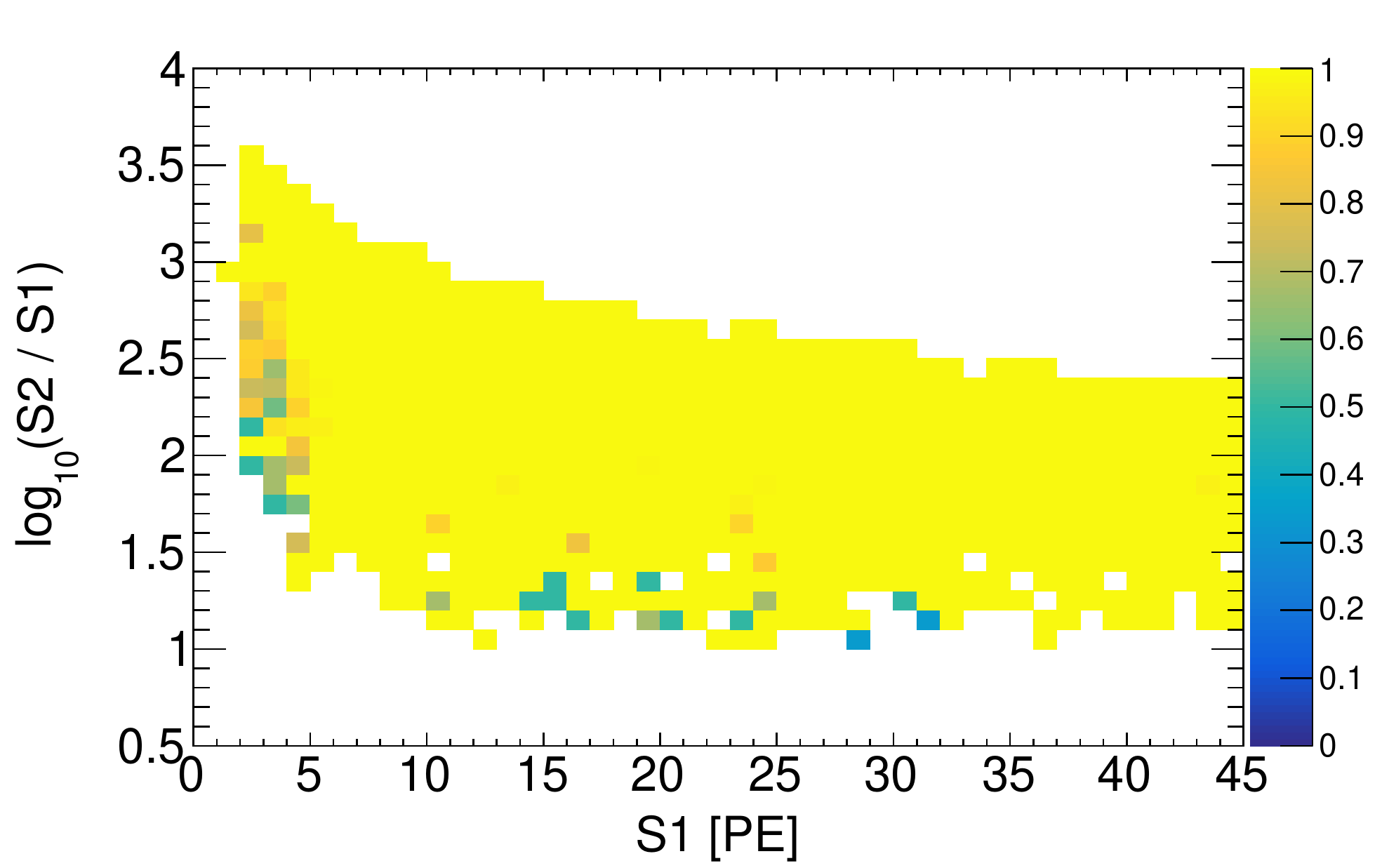}
        \caption{Run 9 (ER)}
    \end{subfigure}
    \begin{subfigure}{0.32\textwidth}
        \includegraphics[width=\textwidth]{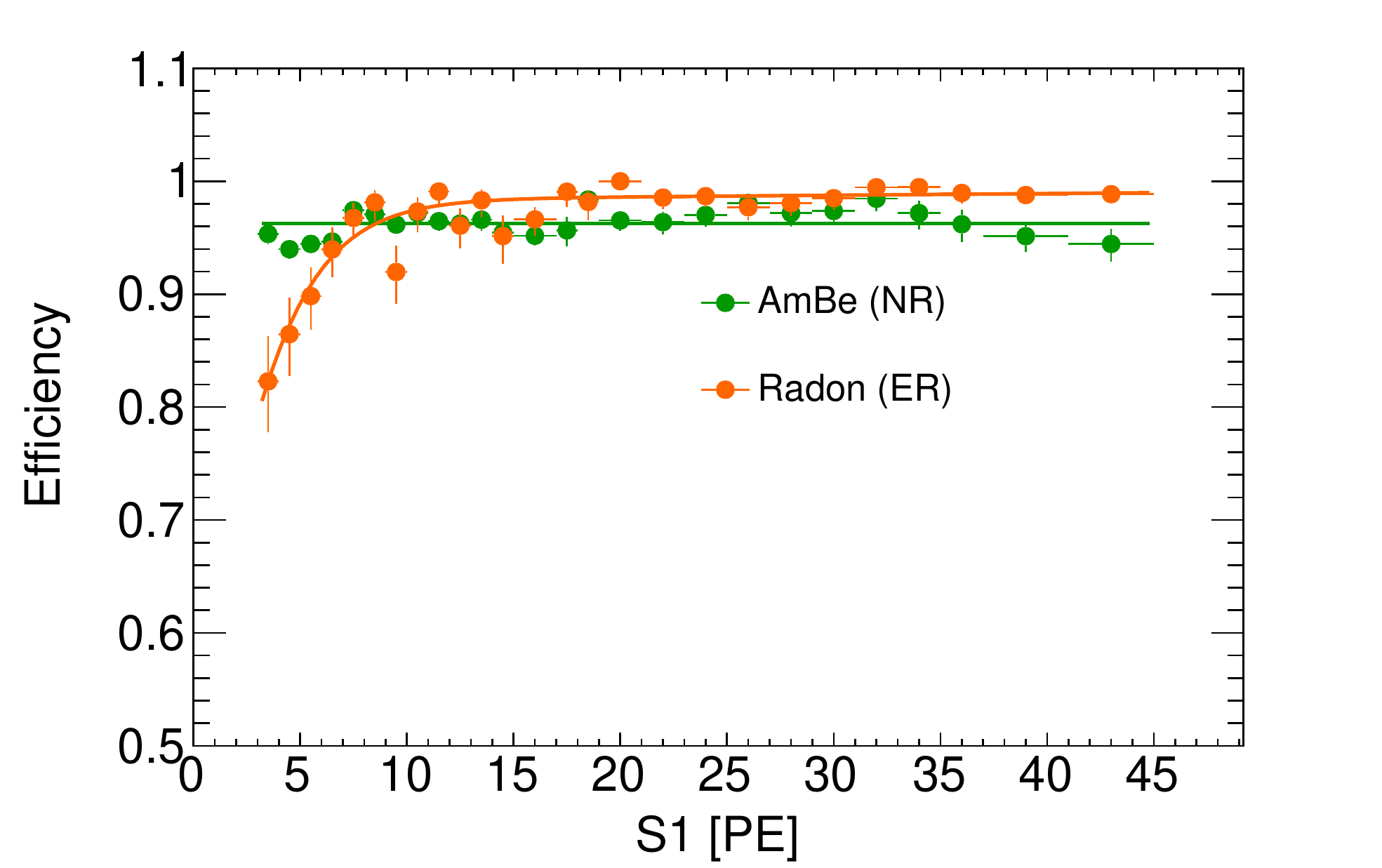}
        \caption{Run 10}
    \end{subfigure}
    \begin{subfigure}{0.32\textwidth}
        \includegraphics[width=\textwidth]{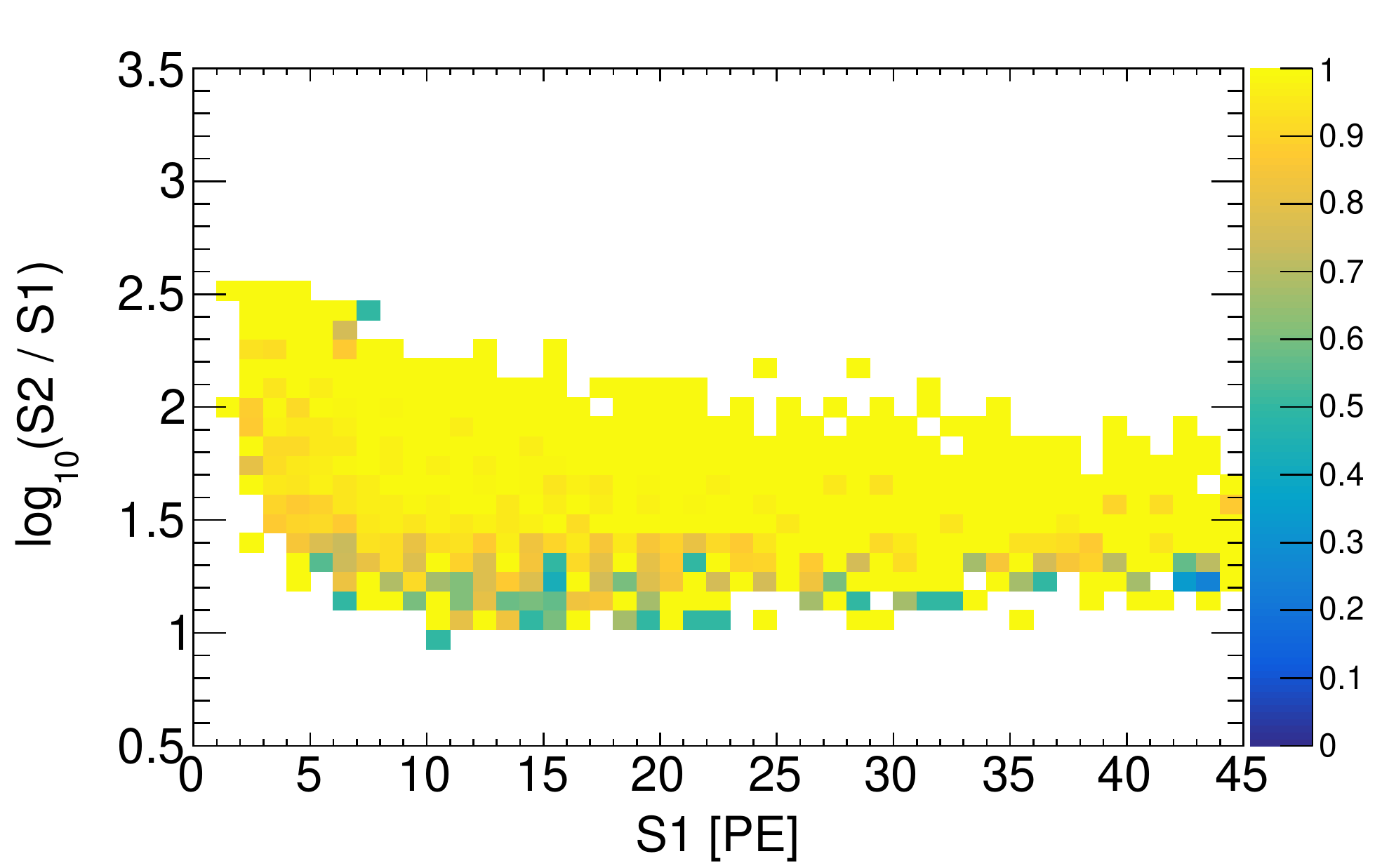}
        \caption{Run 10 (NR)}
    \end{subfigure}
    \begin{subfigure}{0.32\textwidth}
        \includegraphics[width=\textwidth]{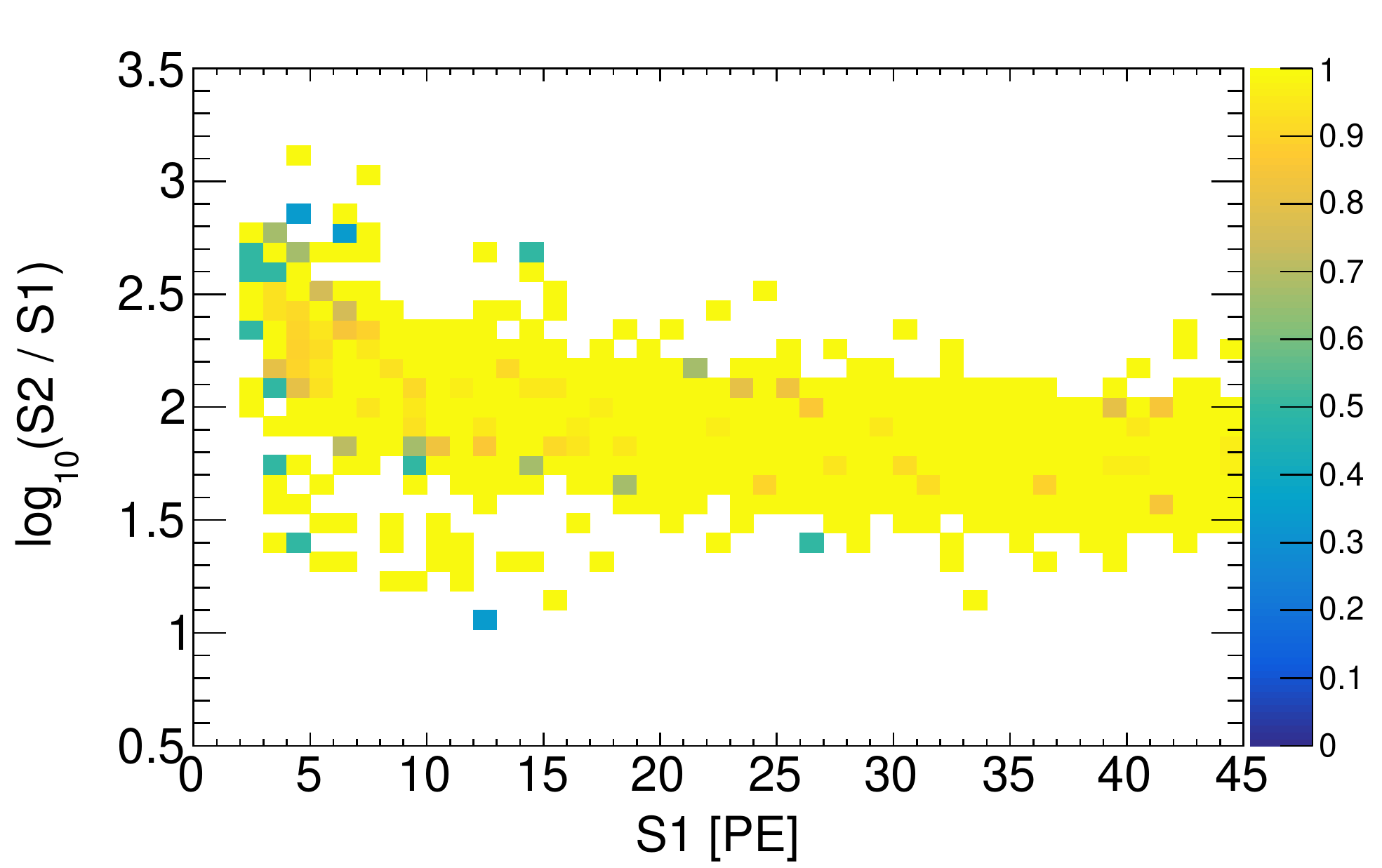}
        \caption{Run 10 (ER)}
    \end{subfigure}
    \begin{subfigure}{0.32\textwidth}
        \includegraphics[width=\textwidth]{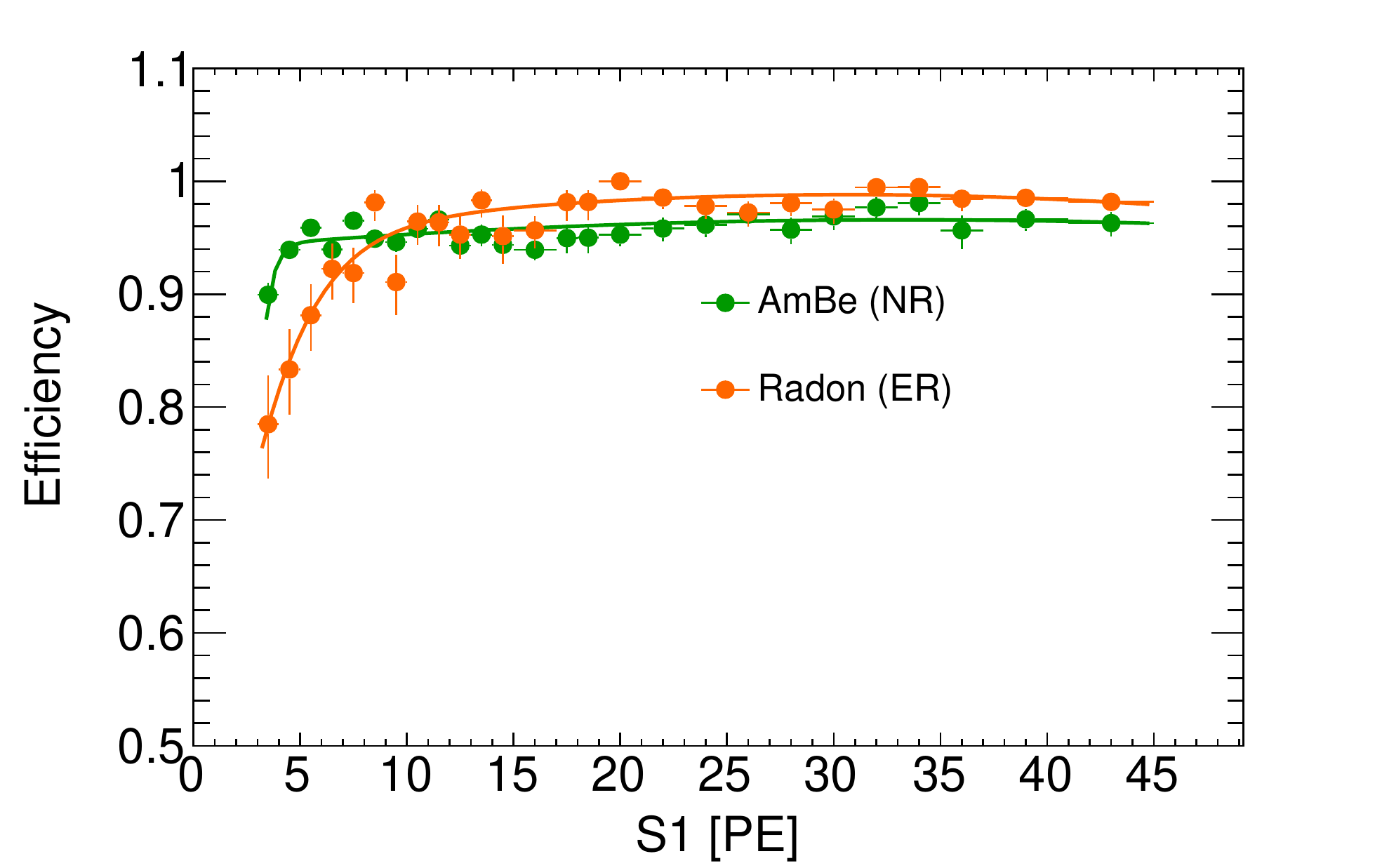}
        \caption{Run 11}
    \end{subfigure}
    \begin{subfigure}{0.32\textwidth}
        \includegraphics[width=\textwidth]{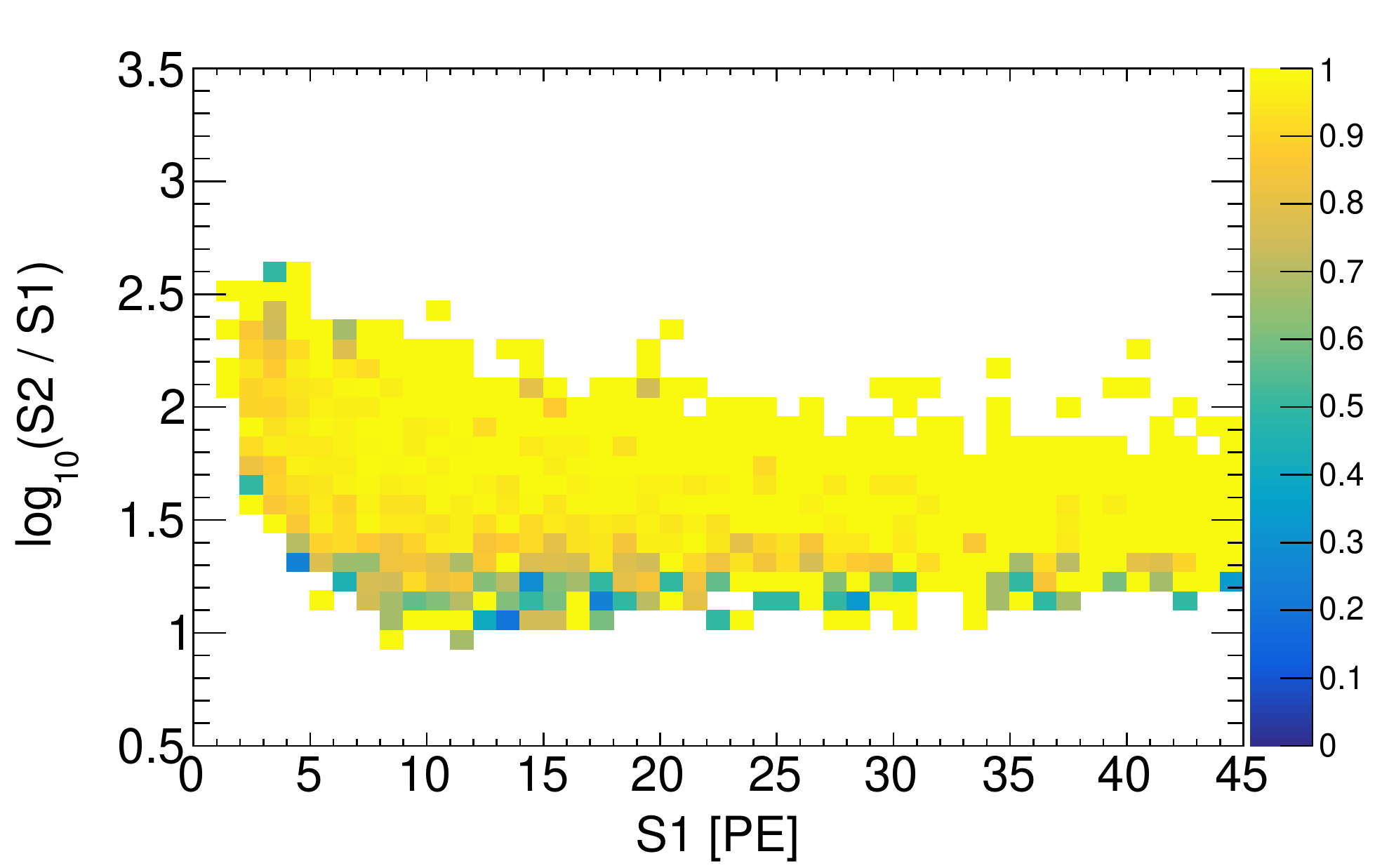}
        \caption{Run 11 (NR)}
    \end{subfigure}
    \begin{subfigure}{0.32\textwidth}
        \includegraphics[width=\textwidth]{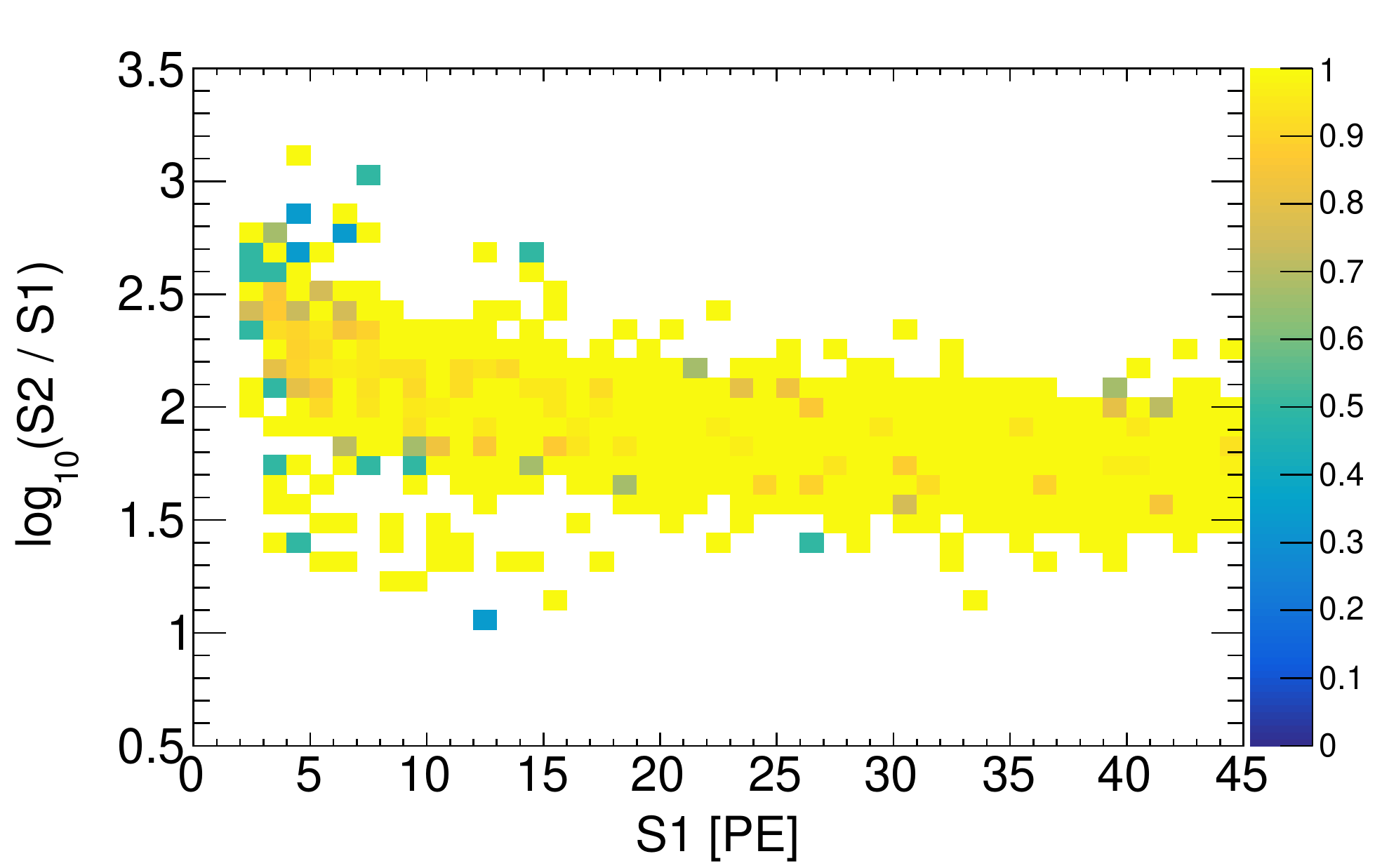}
        \caption{Run 11 (ER)}
    \end{subfigure}    
    \caption{The BDT cut efficiency curves as a function of $S1$ and
      efficiency maps on the $\log_{10}(S2/S1)$ versus $S1$ for
      different calibration data in the dark matter search window for
      different Runs.}
    \label{fig:data_efficiency}
\end{figure}

In the analysis, the BDT cut is not only applied to the events below
the NR median, but applied to all the events in the search window. The
efficiencies of BDT to different types of events are extracted by
using the calibration data sets, shown in
Figure~\ref{fig:data_efficiency}. The BDT cut efficiencies for the ER
and NR calibration data expressed as functions of $S1$, are used to
build the final signal model~\cite{PandaX-II:2021jmq}.  The
efficiencies for ER events are lower than those of NR events when
$S1<8$~PE, in all of the data set. From the 2D efficiency maps, it is
observed that in the region of low $S1$, the ER events with a higher
ratio of $S2/S1$ are suppressed heavily in Runs 10 and 11. On the
contrary, more ER events with smaller $S2/S1$ in the same region are
suppressed in Run 9. The different distributions of $S2$ related
variables of the different ER calibration data may result in the
different efficiencies.  The distribution of $\log_{10}(S2/S1)$
vs. $S1$ of accidental background after the BDT cut are used directly
in the model.

The expected numbers of accidental background (below NR median) in
PandaX-II full exposure data set after the BDT cuts are $2.09\pm0.25$
($0.39\pm0.05$), $1.03\pm0.05$ ($0.27\pm0.01$) and $2.53\pm0.24$
($0.77\pm0.07$) for Runs 9, 10, and 11, respectively. The total number
of expected accidental background events below NR median is smaller
than 1.5. Considering that the total data taking period of PandaX-II
is 244.2 days, we have successfully suppressed the accidental
background to a trivial level and improved the final sensitivity for
dark matter search~\cite{Wang:2020coa}.

\section{Summary and Outlook}
\label{sec:summary}
The accidental background is an important composition of the
backgrounds in the dark matter search experiments with dual phase
xenon detector. We discussed the possible origins of the two
components, isolated $S1$ and $S2$, and developed methods to estimate
the level of accidental background in the PandaX-II experiment. The
BDT algorithm is used to distinguish this non-physical background from
real NR signals below the NR median lines, so that the level of this
background is suppressed greatly.

We found that the rate of isolated $S1$ is much higher in Run 9,
during which the PMTs are running with higher gains than in other
runs. This suggests the coincident combination of hits created by dark
noise contributes a large amount to the isolated $S1$. Thus reducing
the dark noise of PMTs is critical for next generation of
experiments~\cite{hongguang,Mount:2017qzi,XENON:2020kmp}.

The BDT method works well in the suppression of the accidental
background in our study. The analysis framework and suppression method
can be used in the data analysis of the subsequent PandaX-4T
experiment~\cite{PandaX-4T:2021bab}. Because the number of accidental
events is nearly proportional to the operation time, only a few of
them have been produced in the commissioning run. They have been
suppressed to a very low level with the quality cuts and therefore
this method is not used in the first PandaX-4T WIMP search. But
PandaX-4T and other similar experiments will be running much longer
than PandaX-II, our study provides a valuable reference to them.  With
the rapid development of the machine learning methods in recent year,
we may expect the methods of neural networks or some others may
achieve equivalent success in this topic.

\section*{Acknowledgment}
\label{sec:ack}
This project is supported in part by a grant from the Ministry of
Science and Technology of China (No. 2016YFA0400301), grants from
National Science Foundation of China (Nos. 12090060, 12005131,
11905128, 11925502, 11775141), and by Office of Science and
Technology, Shanghai Municipal Government (grant No. 18JC1410200). We
thank supports from Double First Class Plan of the Shanghai Jiao Tong
University. We also thank the sponsorship from the Chinese Academy of
Sciences Center for Excellence in Particle Physics (CCEPP), Hongwen
Foundation in Hong Kong, and Tencent Foundation in China. Finally, we
thank the CJPL administration and the Yalong River Hydropower
Development Company Ltd. for indispensable logistical support and
other help.
\bibliographystyle{JHEP} \bibliography{refs}
\appendix
\renewcommand\thefigure{\thesection.\arabic{figure}}
\section{Complementary plots}
\label{sec:appendix}
\setcounter{figure}{0}
\begin{figure}[hbt]
  \centering
  \includegraphics[width=0.65\textwidth]{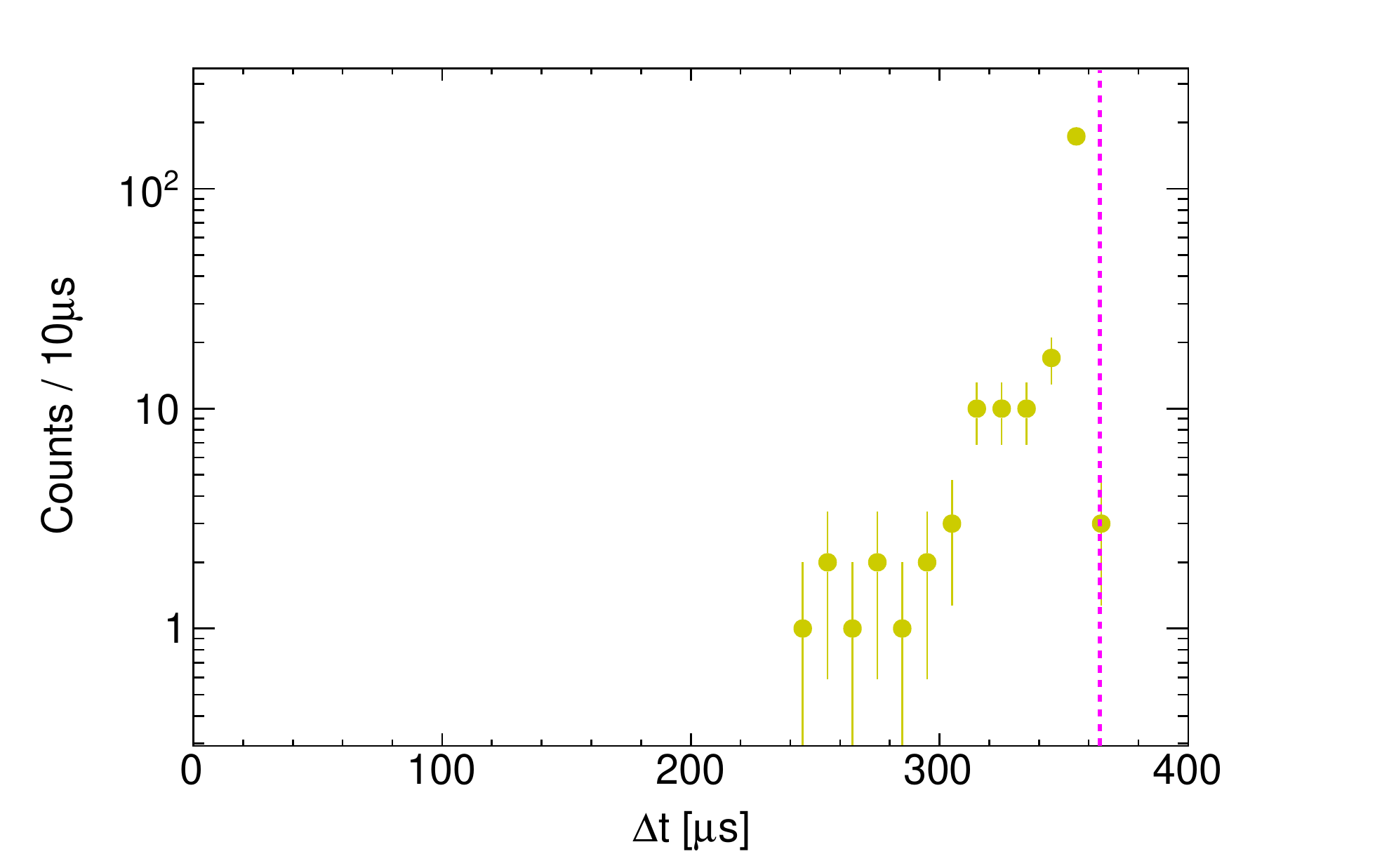}
  \caption{Distribution of the time difference between $S1_{max}$ and
    $S2_{max}$ at the condition where time difference between the
    isolated $S1$ and $S1_{max}$ is smaller than 120~$\mu$s
    ($\Delta t < 120$~$\mu$s) in method 3. The pink dashed line represents
    the maximum drift time.}
  \label{fig:m3_cathode}
\end{figure}

\begin{figure}[hbt]
  \centering
  \includegraphics[width=0.55\textwidth]{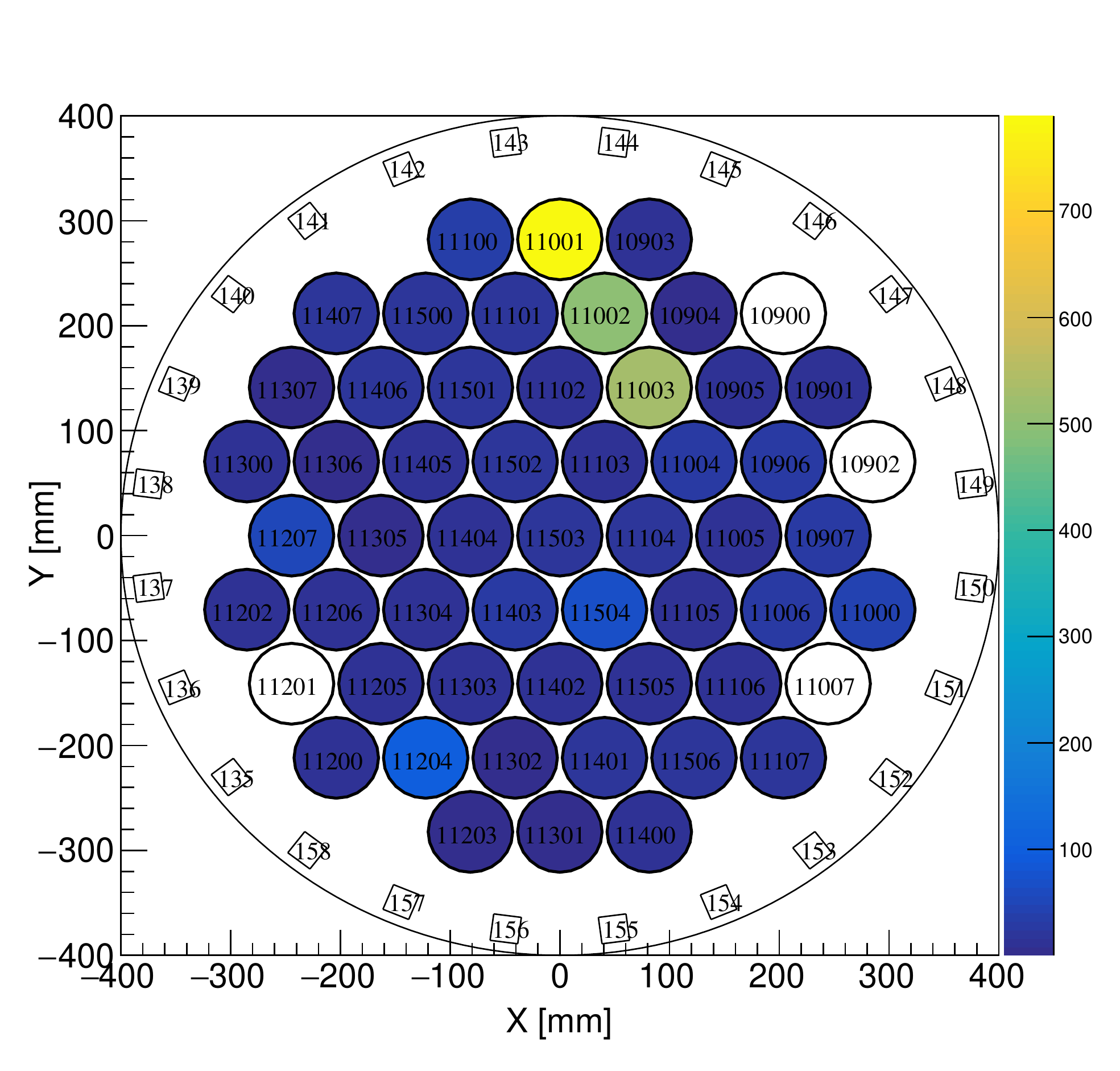}
  \caption{Accumulated charge pattern in the top PMT array of all
    isolated S1s from Mar. 11, 2018 to Apr. 6, 2018. Three PMTs are
    observed to have the largest contribution to these signals. }
  \label{fig:hitpattern_high_rate}
\end{figure}

\begin{figure}[hbt]
  \centering
  \includegraphics[width=0.55\textwidth]{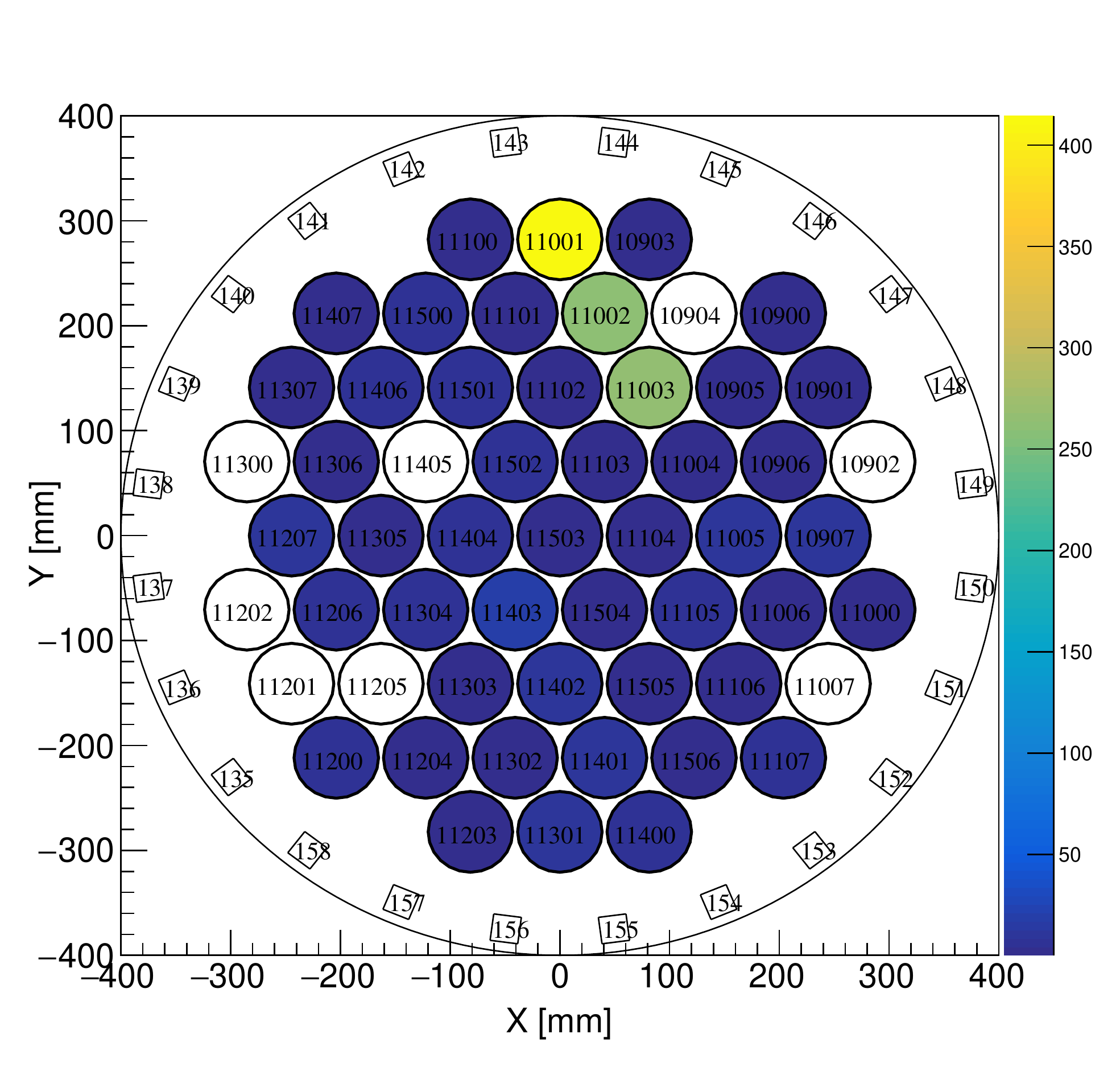}
  \caption{Accumulated charge pattern in the top PMT array of isolated
    S1s in the window of $(10, 12)$~PE in Run 11. Three PMTs are
    observed to have the largest contribution to these signals. }
  \label{fig:hitpattern_10pe}
\end{figure}

\begin{figure}[hbt]
  \centering
  \includegraphics[width=.6\textwidth]{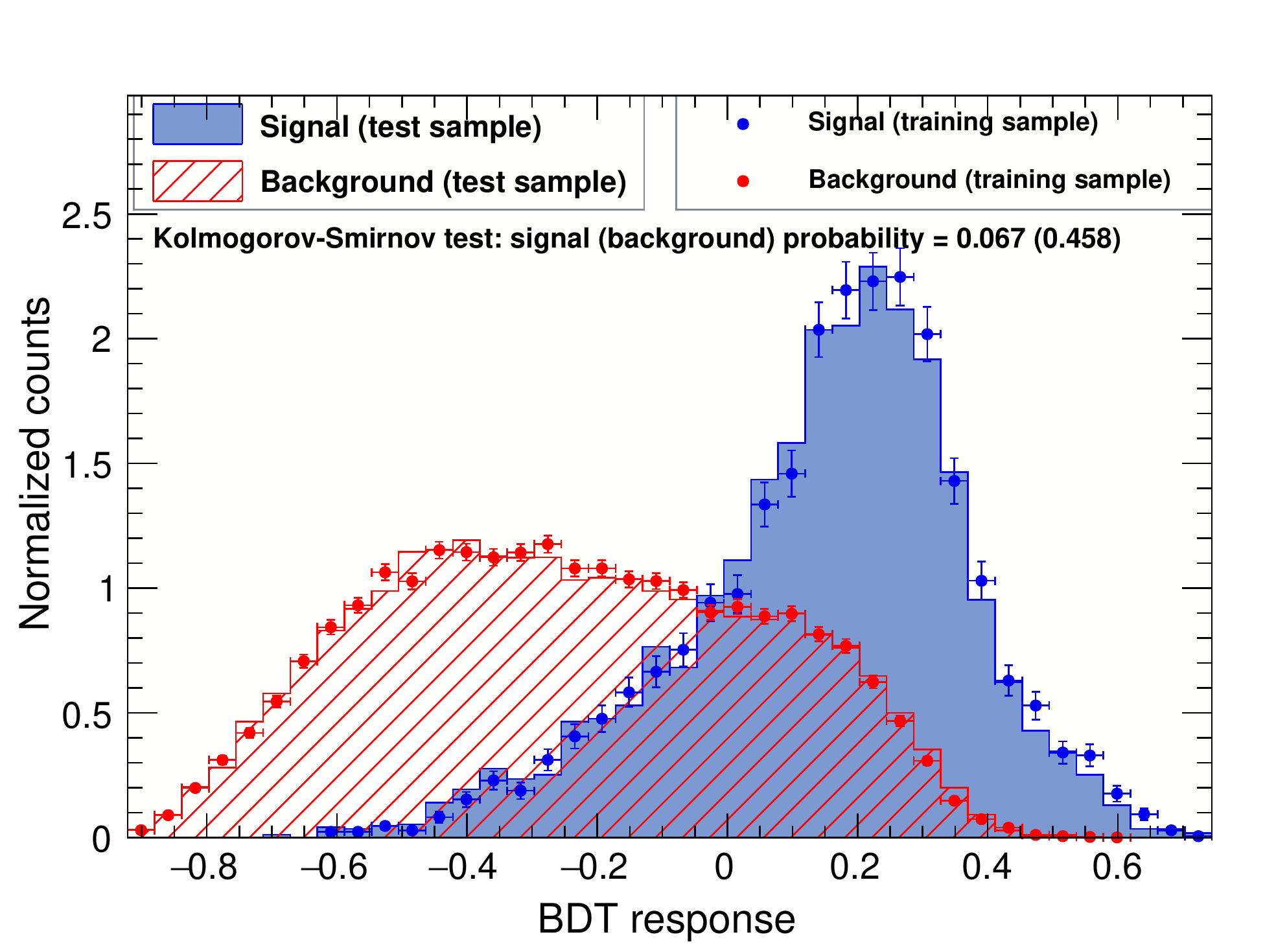}
  \caption{The distributions of BDT response of the train and test
    data samples. The K-S test probabilities are used to indicate the
    overtraining.}
  \label{fig:overtraining_check}
\end{figure}

\begin{figure}[hbt]
  \centering
  \begin{subfigure}{0.40\linewidth}
    \includegraphics[width=\textwidth]{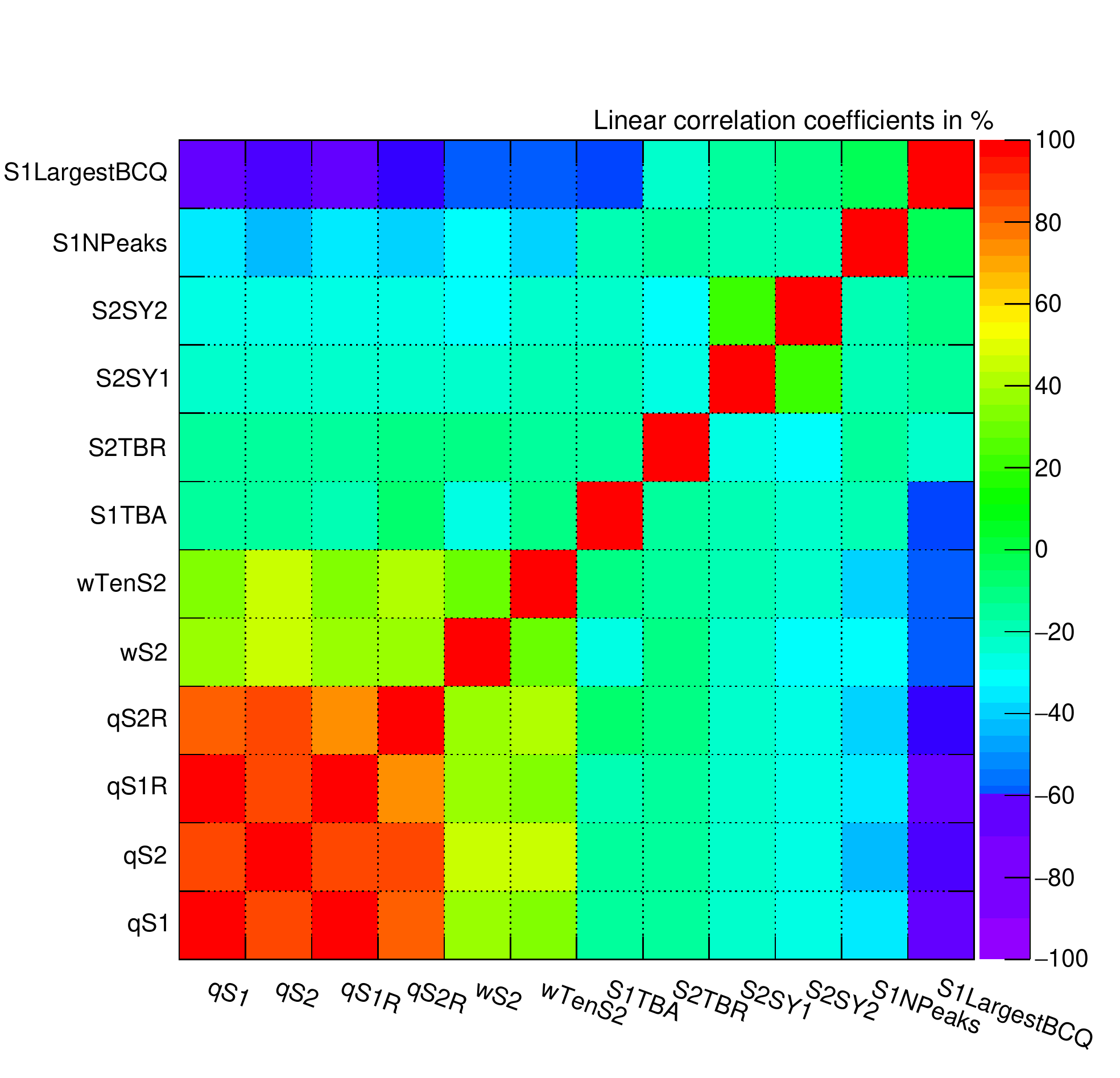}
    \caption{Run 9 (Signal)}
    \label{fig:cor_mat_sig_run9}
  \end{subfigure}
  \begin{subfigure}{0.40\linewidth}
    \includegraphics[width=\textwidth]{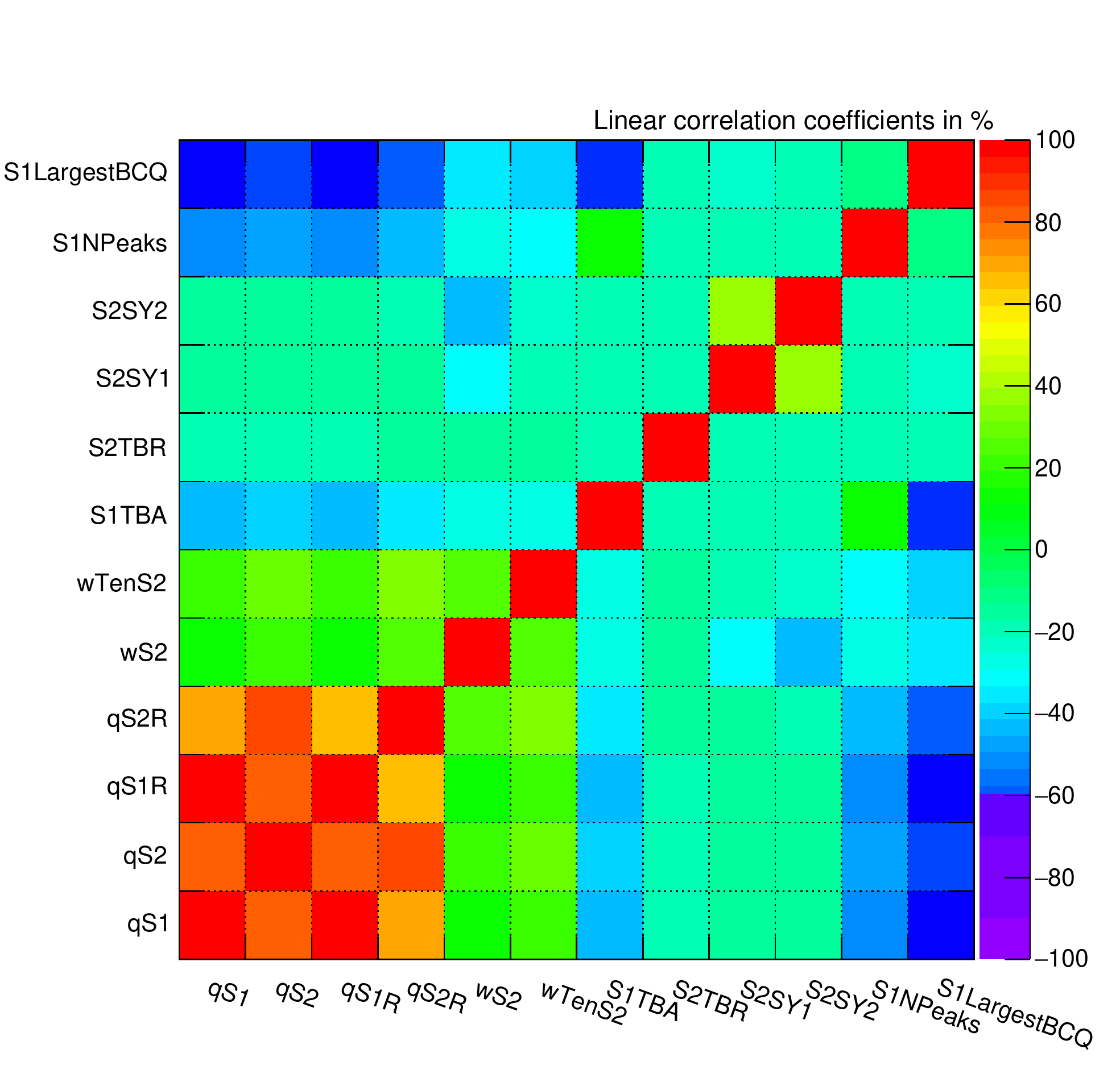}
    \caption{Run 9 (Background)}
    \label{fig:cor_mat_bkg_run9}
  \end{subfigure}
  \begin{subfigure}{0.40\linewidth}
    \includegraphics[width=\textwidth]{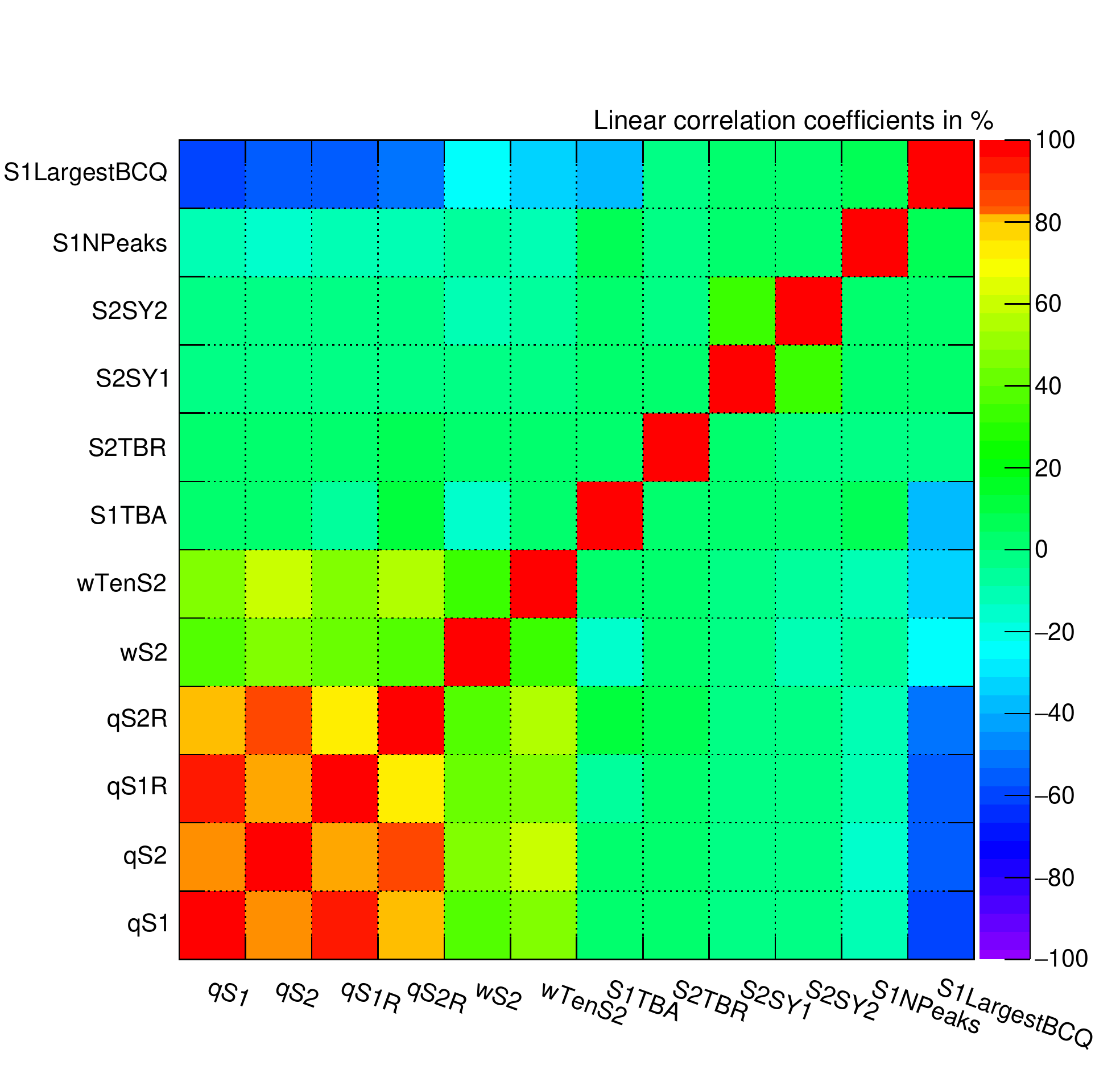}
    \caption{Run 10 (Signal)}
    \label{fig:cor_mat_sig_run10}
  \end{subfigure}
  \begin{subfigure}{0.40\linewidth}
    \includegraphics[width=\textwidth]{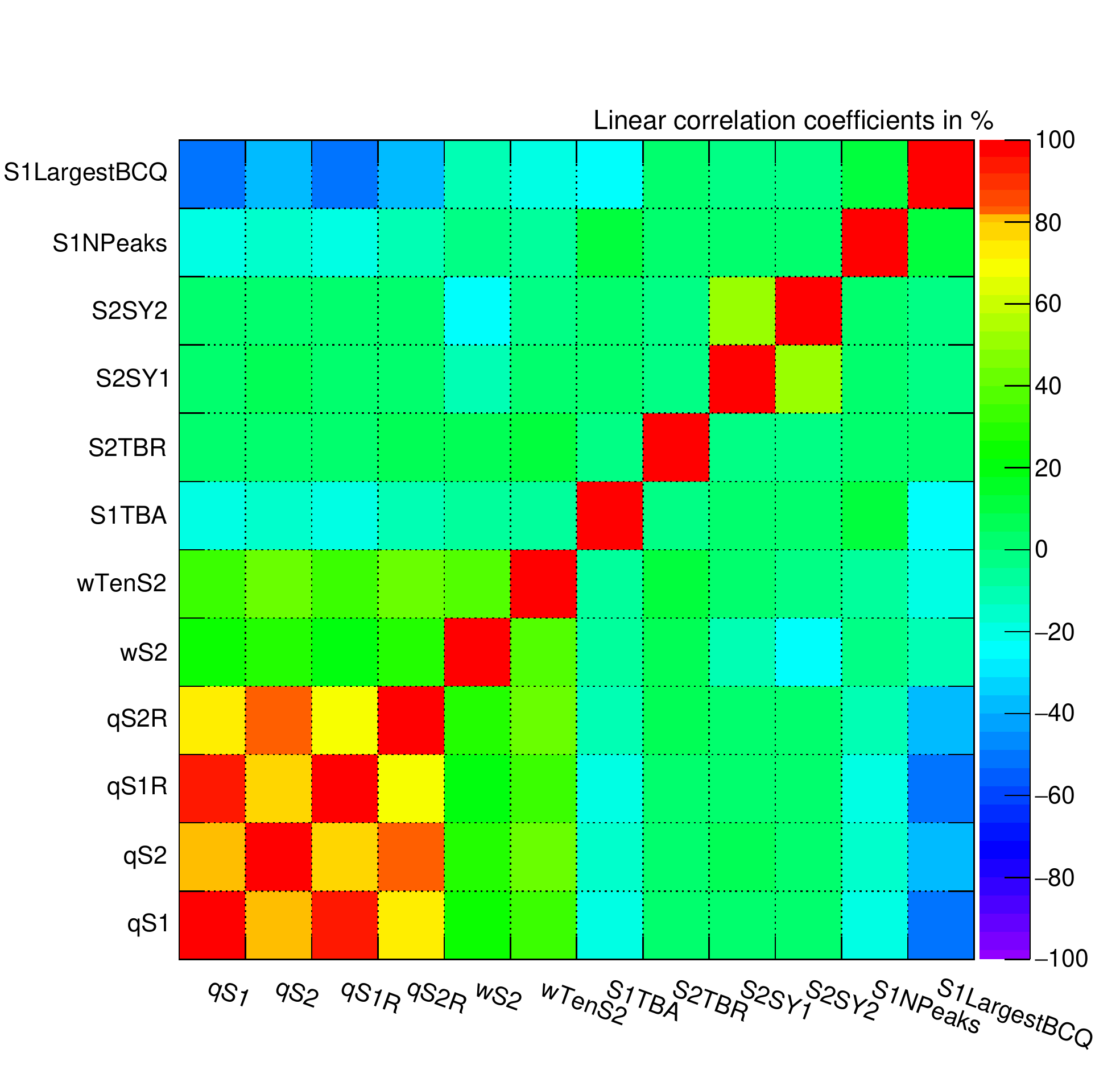}
    \caption{Run 10 (Background)}
    \label{fig:cor_mat_bkg_run10}
  \end{subfigure}
  \begin{subfigure}{0.40\linewidth}
    \includegraphics[width=\textwidth]{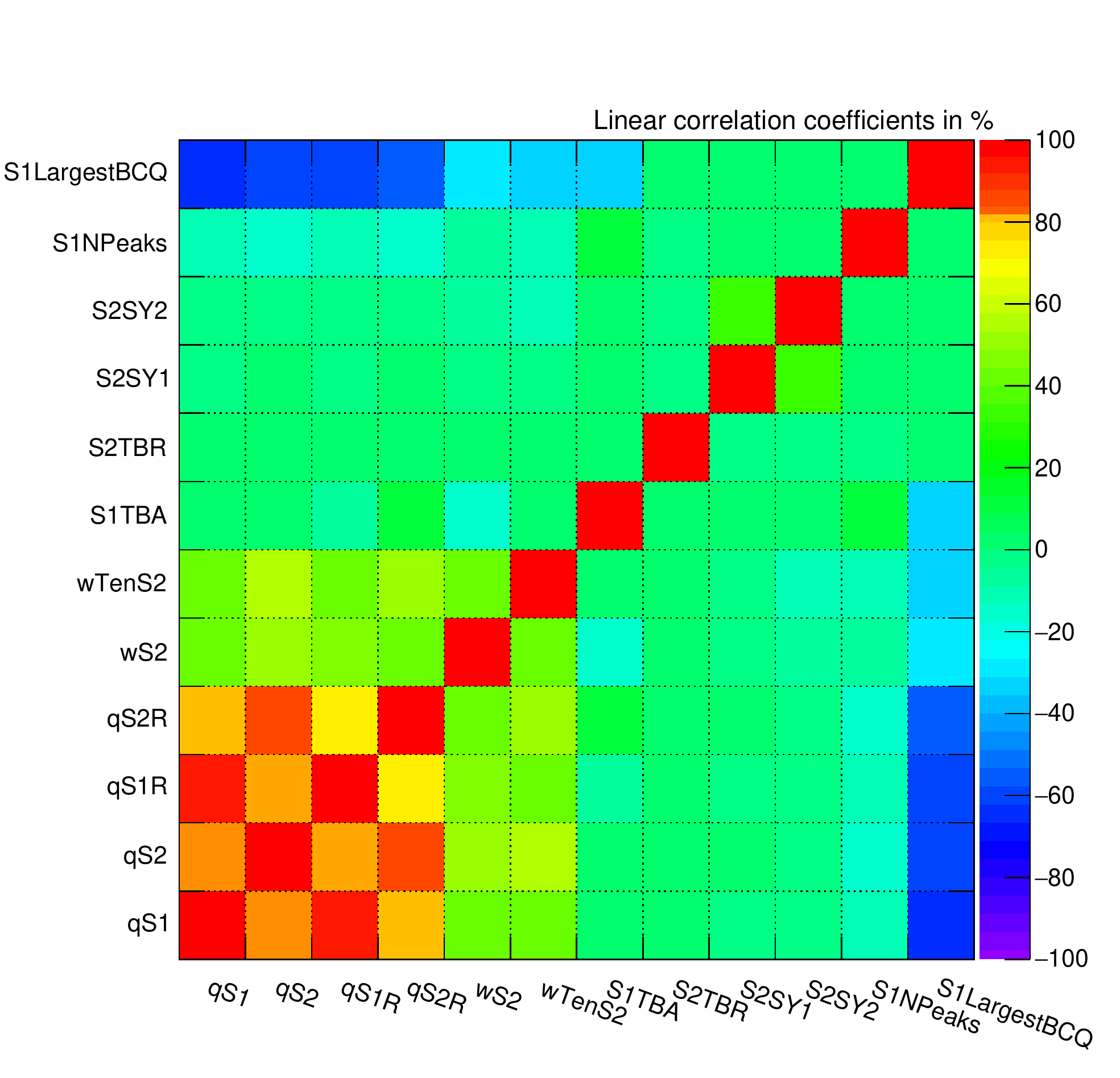}
    \caption{Run 11 (Signal)}
    \label{fig:cor_mat_sig_run11}
  \end{subfigure}
  \begin{subfigure}{0.40\linewidth}
    \includegraphics[width=\textwidth]{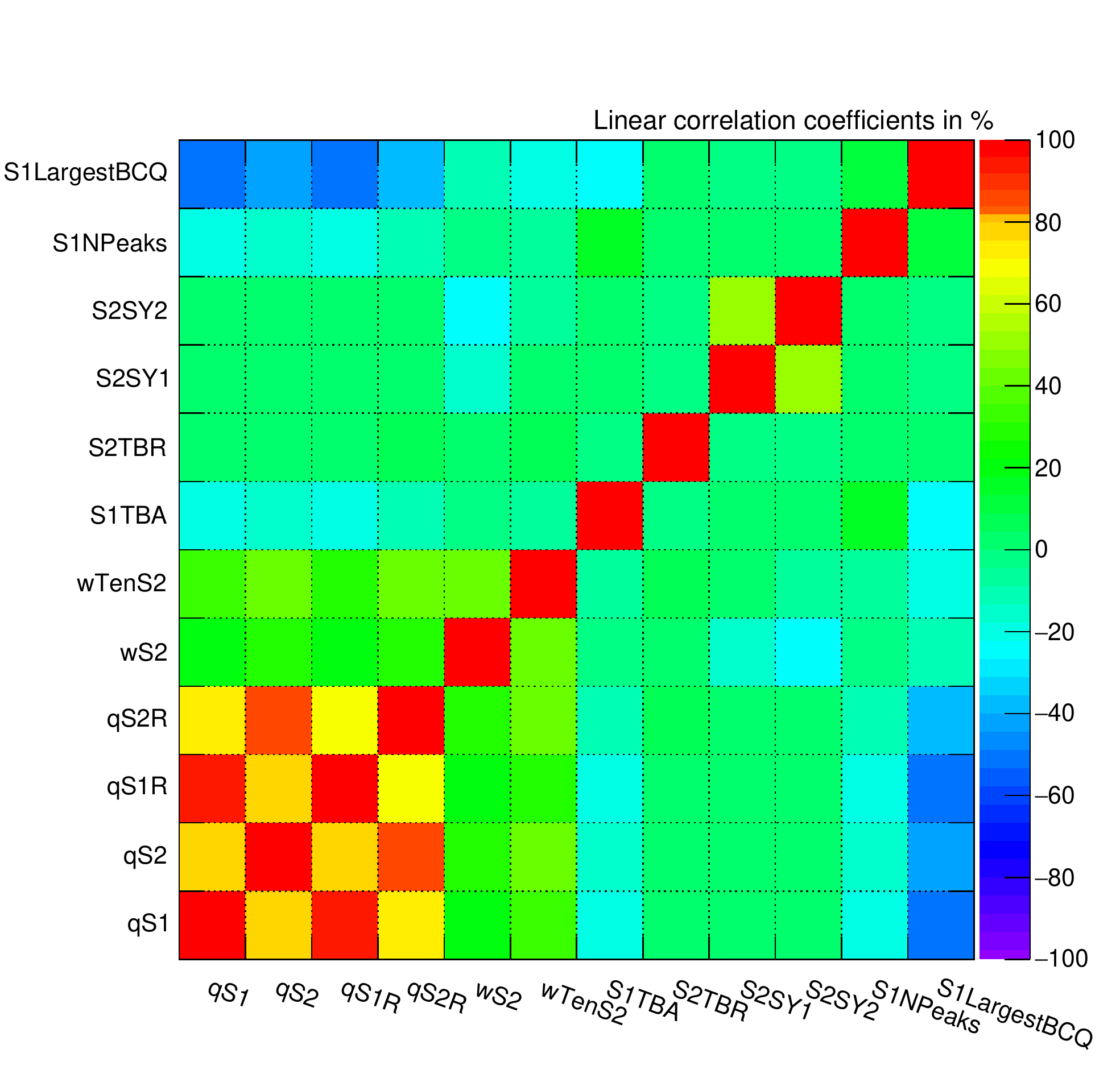}
    \caption{Run 11 (Background)}
    \label{fig:cor_mat_bkg_run11}
  \end{subfigure}
  \caption{Correlations between the variables used for BDT training,
    from the events below the NR median.}
  \label{fig:cor_mat}
\end{figure}
\end{CJK*}
\end{document}